\documentclass[11pt,preprint]{aastex}
\usepackage{epsfig}

\newcommand       \AU           {\,{\rm AU}}          

\newcommand       \cm           {\,{\rm cm}}

\newcommand	  \g		{\,{\rm g}}
\newcommand       \K            {\,{\rm K}}
\newcommand	  \pc		{\,{\rm pc}}

\newcommand	  \yr		{\,{\rm yr}}
\newcommand	  \myr		{\,{\rm Myr}}
\newcommand       \simlt        {\lesssim}
\newcommand       \simgt        {\gtrsim}
\newcommand       \gtsim        {\gtrsim}

\newcommand       \mum          {\,{\rm \mu m}}
\newcommand	  \ppm		{\,{\rm ppm}}
\newcommand	  \Teff	        {T_{\rm eff}}

\newcommand	  \hra	        {{\rm HR\,4796A}}
\newcommand	  \hrb	        {{\rm HR\,4796B}}
\newcommand	  \amin	        {a_{\rm min}}
\newcommand	  \amax	        {a_{\rm max}}
\newcommand	  \rin	        {r_{\rm in}}
\newcommand	  \rout	        {r_{\rm out}}
\newcommand	  \xsun         {\left[{\rm X/H}\right]_{\odot}}
\newcommand	  \csun         {\left[{\rm C/H}\right]_{\odot}}
\newcommand	  \nsun         {\left[{\rm N/H}\right]_{\odot}}
\newcommand	  \osun         {\left[{\rm O/H}\right]_{\odot}}
\newcommand	  \fesun        {\left[{\rm Fe/H}\right]_{\odot}}
\newcommand	  \mgsun        {\left[{\rm Mg/H}\right]_{\odot}}
\newcommand	  \sisun        {\left[{\rm Si/H}\right]_{\odot}}
\newcommand	  \xdust        {\left[{\rm X/H}\right]_{\rm dust}}
\newcommand	  \cdust        {\left[{\rm C/H}\right]_{\rm dust}}
\newcommand	  \ndust        {\left[{\rm N/H}\right]_{\rm dust}}
\newcommand	  \odust        {\left[{\rm O/H}\right]_{\rm dust}}
\newcommand	  \fedust       {\left[{\rm Fe/H}\right]_{\rm dust}}
\newcommand	  \mgdust       {\left[{\rm Mg/H}\right]_{\rm dust}}
\newcommand	  \sidust       {\left[{\rm Si/H}\right]_{\rm dust}}
\newcommand	  \osil         {\left[{\rm O/H}\right]_{\rm sil}}
\newcommand	  \owater       {\left[{\rm O/H}\right]_{\rm water}}
\newcommand	  \xgas         {\left[{\rm X/H}\right]_{\rm gas}}
\newcommand	  \cgas         {\left[{\rm C/H}\right]_{\rm gas}}
\newcommand	  \ngas         {\left[{\rm N/H}\right]_{\rm gas}}
\newcommand	  \ogas         {\left[{\rm O/H}\right]_{\rm gas}}

\newcommand	  \mux          {\mu_{\rm X}}
\newcommand	  \muc          {\mu_{\rm C}}
\newcommand	  \muh          {\mu_{\rm H}}
\newcommand	  \mun          {\mu_{\rm N}}
\newcommand	  \muo          {\mu_{\rm O}}
\newcommand	  \mufe         {\mu_{\rm Fe}}
\newcommand	  \mumg         {\mu_{\rm Mg}}
\newcommand	  \musi         {\mu_{\rm Si}}
\newcommand	  \msil         {m_{\rm sil}} 
\newcommand	  \mcarb        {m_{\rm carb}} 
\newcommand	  \mice         {m_{\rm ice}} 
\newcommand	  \md           {m_{\rm d}}
\newcommand	  \rp           {r_{\rm p}}
\newcommand	  \vsil         {v_{\rm sil}}
\newcommand	  \vcarb        {v_{\rm carb}}
\newcommand	  \vice         {v_{\rm ice}}
\newcommand	  \rhosil       {\rho_{\rm sil}}
\newcommand	  \rhocarb      {\rho_{\rm carb}}
\newcommand	  \rhoice       {\rho_{\rm ice}}
\newcommand	  \Pice         {P^{\prime}}
\newcommand	  \sigmap       {\sigma_{\rm p}}
\newcommand	  \sigmar       {\sigma(r)}
\newcommand{\figwidth}{6.0in}

\countdef\decade=200
\advance\decade by \year
\countdef\hours=201
\advance\hours by \time
\divide\hours by 60
\countdef\mins=202
\advance\mins by \hours
\multiply\mins by 60
\multiply\hours by 100
\countdef\miltime=203
\advance\miltime by \hours
\advance\miltime by \time
\advance\miltime by -\mins

\shorttitle{Infrared Emission from the HR\,4796A Disk}

\begin{document}

\title{
 \vspace*{-2.0em}
  {\normalsize\rm {\it The Astrophysical Journal},
   vol. 590, pp.\,368--378}\\
 \vspace*{1.0em}
Modelling the Infrared Emission from the HR\,4796A Disk 
	 }

\author{Aigen Li and J.I. Lunine}
\affil{Theoretical Astrophysics Program,
       Departments of Astronomy and Planetary Sciences,
       University of Arizona, Tucson, AZ 85721;\\
        {\sf agli@lpl.arizona.edu, jlunine@lpl.arizona.edu}}

\begin{abstract}
We model the spectral energy distribution (SED) from the mid-infrared 
to submillimeter of the ring-like disk of $\hra$, the dustiest A-type 
star. We consider dust made either of coagulated but otherwise unaltered 
protostellar interstellar grains, or grains that are highly-processed 
in a protostellar/protoplanetary nebula with silicate dust annealed 
and carbon dust oxidized. Both types of dust are successful in reproducing
the observed SED, provided that the grains are highly fluffy, with a vacuum
volume fraction of $\sim 90\%$. We find no evidence for the existence of 
a hot ``zodiacal dust'' component a few AU from the star, which was
suggested by previous workers to account for the 10$\mum$-wavelength 
emission. 
\end{abstract}


\section{Introduction\label{sec:intro}}
$\hra$ is a nearby (distance to the Earth $d\approx 67\pm 3\pc$) 
young main-sequence (MS) star (age $\approx 8\pm 3\myr$) of 
spectral type A0\,V (effective temperature $\Teff\approx 9500\K$).
The $\hra$ dust disk has recently aroused considerable interest because
(1) it has the largest fractional infrared (IR) luminosity relative to 
the total stellar luminosity ($L_{\rm IR}/L_{\star}\approx 5\times 10^{-3}$; 
$L_\star \approx 21\,L_\odot$)
among the $\sim 1500$ A-type MS stars in the {\it Bright Star Catalogue} 
(Jura 1991); (2) unlike the majority of circumstellar disks, the $\hra$
disk displays a ring-like structure peaking at $\sim 70\AU$ from 
the central star and abruptly truncated both interior and outside 
with a width of $\simlt 17\AU$ (Schneider et al.\ 1999);
(3) its young-age nature places it at a somewhat transitional 
stage between massive gaseous protostellar disks around young 
pre-MS T-Tauri and Herbig Ae/Be stars ($\sim 1\myr$) 
and much evolved and tenuous debris disks around MS ``vega-type''
stars ($\sim 100\myr$).

The physical and chemical properties of the dust in the $\hra$ disk,
which play a significant role in understanding the creation, dynamical 
evolution and structural properties of the disk, are poorly constrained: 
(1) Based on the lifetime of dust against loss by Poynting-Robertson drag,
Jura et al.\ (1993) suggested that the $\hra$ dust is larger than $10\mum$ 
in radius and the dust is a remnant of the protostellar nebula;
(2) From an analysis of radiation pressure vs. gravitational attraction,
Jura et al.\ (1995) found that the minimum radius of grains orbiting
$\hra$ is about $3\mum$ and argued that these grains have undergone 
measurable coalescence; 
(3) In order to explain the $\simgt 40\AU$ inner hole (Jura et al.\ 1995) 
and the $\approx 110\K$ black-body approximation for the 10--100$\mum$ 
dust emission, Jura et al.\ (1998) proposed that the $\hra$ grains are
largely composed of ice particles with a typical radius of $\sim 100\mum$ 
which originate from a protocometary cloud;
(4) In order to explain the entire 12.5$\mum$ emission and 
the 20.8$\mum$ residual emission unaccounted for by their simple 
model, Koerner et al.\ (1998) proposed the existence of a tenuous dust
component at a distance of 3--6$\AU$ from the star which is heated to 
about 200--300$\K$, similar to the zodiacal dust in our own solar system; 
(5) From the red reflectance of the disk, Schneider et al.\ (1999) 
argued that the mean dust size must be larger than several microns 
and the dust is circumstellar debris rather than interstellar in origin;
(6) Based on a detailed modelling of the spectral energy distribution
(SED) of the $\hra$ disk, Augereau et al.\ (1999) proposed a two-component
model consisting of a cold annulus peaking at $70\AU$ from the star
(made of interstellar dust-type grains ranging from $10\mum$ to 
a few meters) and a hot population at 9--10$\AU$ from the star 
(made of comet-like grains of radii $\approx 450\mum$);
(7) Using the flux ratio at 10.8$\mum$ and 18.2$\mum$ 
and assuming ``astronomical silicates'', 
Telesco et al.\ (2000) inferred a ``characteristic'' diameter
of $\approx 2-3\mum$ and argued that these grains are unlikely to be
primordial; instead, they are probably products of recent collisions
of large bodies. 

It is the purpose of this work to constrain the dust properties
(size, composition, and morphology) of the $\hra$ disk.
In general, in a dusty system, dust spatial distribution and 
dust sizes cannot be uniquely determined by the SED {\it alone}.
Given that the distribution of the dust in the $\hra$ disk is well 
constrained by the near-IR imaging of scattered starlight 
(Schneider et al.\ 1999) and, to a less degree, by the mid-IR imaging 
of dust thermal emission (Jayawardhana et al.\ 1998; Koerner et al.\ 1998; 
Telesco et al.\ 2000), an attempt is therefore made in this work to 
infer the $\hra$ dust properties by modelling the full spectral energy 
distribution from the mid-IR to the submillimeter wavelengths. 

Lacking a priori knowledge of the composition of the dust in
the $\hra$ disk, we consider two extreme dust types: 
(1) ``cold-coagulation'' dust -- dust in protoplanetary disks is 
formed through cold aggregation of unaltered interstellar materials
-- this is the case in the outer parts of the disk where interstellar
dust originating from the parent molecular cloud (out of which the disk
forms) is not likely to have significantly changed its composition 
during passage through the weak shock front when the disk is first 
created, except volatile ice mantles around the refractory cores 
may partly sublimate, but the recondensation of volatiles occurs 
efficiently behind the shock (see Beckwith, Henning, \& Nakagawa 
2000 for a review); 
(2) ``hot-nebula'' dust -- dust has undergone significant destruction 
and modification in protoplanetary accretion disks through annealing 
of amorphous silicates and oxidation of carbonaceous dust in the warm 
inner regions of the accretion disk (Gail 2001); 
turbulent radial mixing (Gail 2001) 
and/or outflows driven by ``X-winds'' (Shu, Shang, \& Lee 1996) 
then carries these heavily processed dust into cold regions of 
the disk where it is mixed with freshly accreted material 
from the parent molecular cloud;
as an extreme, we consider a dust model in which the carbonaceous 
dust component has been fully destroyed by oxidation so that only 
crystalline silicate dust remains. 
In reality, the dust in protoplanetary disks would undoubtedly 
be intermediate between heavily processed dust and 
aggregates of unaltered interstellar dust.
Since planetesimals and cometesimals are formed by coagulation
of such dust aggregates, the dust generated by collisions of 
planetesimals and cometesimals should resemble the original dust
from which they are built up.

We first discuss in \S\ref{sec:method} the general constraints on 
the $\hra$ dust and disk properties.
We then model the observed SED in \S\ref{sec:cold} and \S\ref{sec:hot} 
in terms of the ``cold-aggregation'' model of unaltered interstellar 
dust and the ``hot-nebula'' model with all amorphous silicate dust 
annealed and all carbonaceous dust oxidized, respectively.
We find that, for both models, (1) highly porous grains with a porosity
$P\approx 0.90$ provide excellent fits to the entire SED; 
(2) there appears to be no need for a hot ``zodiacal cloud'' dust 
component (\S\ref{sec:zodi}) which was previously suggested 
by Koerner et al.\ (1998) and Augereau et al.\ (1999).  
Therefore, the presence of either type of particles 
in the disk is possible. We show in \S\ref{sec:pwl} that,
although good fits to the SED can be achieved by models with 
a single power-law dust spatial distribution, they are apparently
in conflict with the imaging observations of scattered starlight 
and dust thermal emission.
In \S\ref{sec:sirtf} we calculate the dust IR intensities
integrated over the SIRTF/IRAC and MIPS bands predicted for
our best-fitting models.  
The major conclusions are summarized in \S\ref{sec:summary}.

\section{Modelling the Dust IR Emission\label{sec:model}}
A grain in the optically thin dust disk of $\hra$ absorbs stellar 
ultraviolet/optical photons and then re-radiates the energy in 
the IR. To calculate the disk's emission spectrum, knowledge of 
its morphology, composition, and size is required.  

\subsection{General Considerations on Dust and Disk 
Properties\label{sec:method}}
A fluffy structure is expected for dust in protoplanetary disks
as a result of its coagulational growth process. We characterize 
its fluffiness by porosity $P$, the fractional volume of vacuum. 
We assume all grains are spherical in shape. 

In \S\ref{sec:composition} we estimate that the ``cold-coagulation''
model leads to porous dust consisting of amorphous silicate and 
carbonaceous materials (and H$_2$O-dominated ices in regions colder
than $\sim 110-120\K$) with a mixing ratio 
of $\mcarb/\msil \approx 0.7$ 
[and $\mice/(\msil+\mcarb) \approx 0.8$ for cold regions].
For the ``hot-nebula'' dust model, we assume that 
the dust is exclusively composed of crystalline silicate dust 
(and ices in cold regions).

We assume a power law dust size distribution 
$dn(a)/da \propto a^{-\alpha}$ which is characterized by 
a lower-cutoff $\amin$, upper-cutoff $\amax$ 
and power-law index $\alpha$ (where $a$ is the spherical radius).
We take $\amin=1\mum$, the smallest value among previous estimations
of $\amin$ (see \S\ref{sec:intro});\footnote{%
 A fluffy grain of $a=1\mum$ with a porosity of $P=0.90$
 consists of $\sim 100$ constituent individual (interstellar) particles
 which have a typical size of $a\sim 0.1\mum$ (see Li \& Greenberg 1997).
 Models with a smaller $\amin$ ($=0.1\mum$) and 
 a larger $\amin$ ($=10\mum$) will be discussed in \S\ref{sec:pwl} 
 }
and $\amax=1\cm$ (this is not a critical parameter since grains 
larger than $\sim 100\mum$ are like blackbodies
and their IR emission spectra are size-insensitive).

The dust spatial distribution is well constrained by 
the near-IR imaging of scattered starlight (Schneider et al.\ 1999)
to be a sharply-truncated ring-like structure peaking at a radial 
distance of $\sim 70\AU$ from the star with a characteristic width 
of $\simlt 17\AU$. Therefore, we adopt a Gaussian function
for the dust spatial density distribution:\footnote{%
  Following Kenyon et al.\ (1999), we assume an exponential, 
  radial-independent, vertical distribution 
  $dn/dr \propto \exp[-\left(z/\sqrt{2}H\right)^2]$
  where the vertical scale height $H$ is $\approx 0.5\AU$.
  Alternatively, $H$ can be determined from the
  vertical hydrostatic equilibrium assumption
  ($H\propto r^{3/2}$; see Appendix B in 
  Li \& Lunine 2003). 
  But the knowledge of the dust vertical distribution 
  is not required and it does not affect our results 
  since in modelling the SED of the $\hra$ disk, 
  what actually involves is
  the dust surface density distribution $\sigmar$.  
  } 
$dn(r)/dr \propto \exp[-4\ln2\{(r-\rp)/\Delta\}^2]$
which is characterized by the radial position $\rp$ 
where $dn(r)/dr$ peaks and the full width half maximum (FWHM) $\Delta$.
The dust surface density distribution obtained by
integrating $dn/dr$ over the perpendicular (to the disk plane)
path lengths can be expressed as
$\sigmar = \sigmap \exp[-4\ln2\{(r-\rp)/\Delta\}^2]$
where $\sigmap$ is the mid-plane ($z=0$) surface density at $r=\rp$.
This distribution function, with $\rp=70\AU$ and $\Delta=12\AU$, 
was successful in modelling the scattered-light images 
(Kenyon et al.\ 1999).
A Gaussian-type dust distribution was also adopted by
Klahr \& Lin (2000) to model the dynamics of the dust ring
around $\hra$.
We fix $\rp$ to be at $\rp=70\AU$ and take $\Delta=15\AU$, 
the mean value of the determinations of
Schneider et al.\ (1999; $\Delta\simlt 17\AU$)
and Kenyon et al.\ (1999; $\Delta=12\AU$).  
The inner boundary $\rin$ is set to be where grains are heated to 
$\gtsim 1500\K$, and hence, $\rin$ is a function of dust size.
For micron-sized grains, the inner boundary is roughly $\rin=0.15\AU$. 
The outer boundary is taken to be $\rout=250\AU$ which is 
expected from the disk truncation caused by the tidal effects 
of $\hrb$, a companion star of $\hra$ (Jayawardhana et al.\ 1998).  

Therefore, we are only left with two free parameters:
(1) the dust porosity $P$, and
(2) the dust size distribution power index $\alpha$.

Dielectric functions are taken from 
(1) Draine \& Lee (1984) for amorphous silicate dust;
(2) Li \& Draine (2001a) for crystalline silicate dust;
(3) Li \& Greenberg (1997) for carbonaceous dust;
(4) Li \& Greenberg (1998) for H$_2$O-dominated ice.
The Bruggman effective medium theory (Bohren \& Huffman 1983)
is used to calculate the mean dielectric functions for the fluffy 
heterogeneous dust aggregates. 
Absorption cross sections are obtained using Mie theory.
Approximating the $\hra$ radiation field by the Kurucz model 
atmosphere spectrum for A0\,V stars (Kurucz 1979),
dust equilibrium temperatures are then derived 
by balancing absorption and emission. 
For a given dust size distribution and a given disk structure (dust 
spatial density distribution), the emergent IR emission spectrum can 
be obtained by integrating over the dust size range, 
and over the entire disk. The calculated IR spectrum is then
compared with the available photometric data for the $\hra$ disk
compiled by Augereau et al.\ (1999).

\subsection{``Cold-Coagulation'' Dust Model\label{sec:cold}}
We first model the $\hra$ SED in terms of the dust generated 
from collisions of planetesimals and cometesimals formed in 
the disk (see \S\ref{sec:rppr}) through cold-coagulation of 
unaltered interstellar grains from its parent molecular cloud. 
Using the dust composition discussed in \S\ref{sec:composition}, 
we calculate the model IR spectra for a wide range of dust 
porosities $P$ and a wide range of dust size distribution 
power-indices $\alpha$ to search for good fits. 
We note that $P$ refers to the porosity of
refractory dust -- $P$ will be reduced to $\Pice$ for dust
with ices (see \S\ref{sec:porosity}).

\begin{figure}[h]
\begin{center}
\epsfig{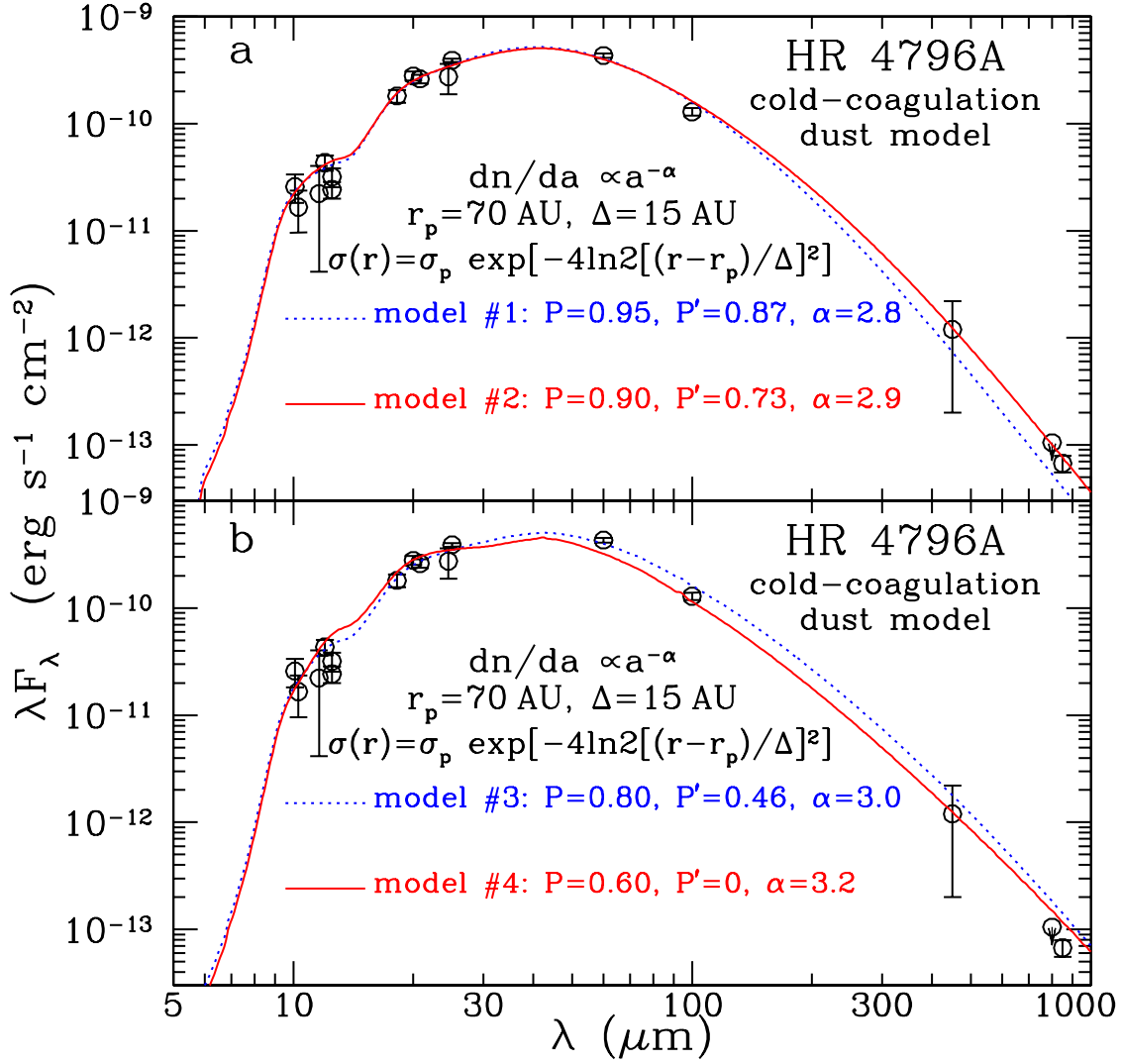}
\end{center}\vspace*{-1em}
\caption{
        \label{fig:cold}
        \footnotesize
        Comparison of the observed spectral energy distribution
        of the $\hra$ dust disk to theoretical IR emission spectra
        calculated from the ``cold-coagulation'' model (porous aggregates 
        of amorphous silicate dust and carbonaceous dust plus ices
        in cold regions). The dust in this ring-like disk is taken 
        to have a Gaussian distribution function
        $\sigmar = \sigmap \exp[-4\ln2\{(r-r_0)/\Delta\}^2]$ 
        with its peak at $r_{\rm p}=70\AU$ and a FWHM $\Delta=15\AU$
        ($r\in [0.15\AU,250\AU]$) as inferred from the scattered light
        images (Schneider et al.\ 1999; Kenyon et al.\ 1999).
        We assume a power-law dust size distribution 
        $dn(a)/da \propto a^{-\alpha}$ ($a\in [1\mum,1\cm]$).
        Upper panel (a): dotted line (model no.\,1) -- 
            best-fit $P=0.95$ ($\Pice\approx 0.87$; 
            see \S\ref{sec:porosity}) model 
            ($\alpha \approx 2.8$, $\md \approx 2.67\times 10^{27}\g$,
             $\sigmap\approx 2.97\times 10^4\cm^{-2}$, 
            $\chi^2/N\approx 2.85$ where $N=18$ [16 data points plus 
            2 free parameters $P$ and $\alpha$; see text]);
            solid line (model no.\,2) -- 
            best-fit $P=0.90$ ($\Pice\approx 0.73$) model
            ($\alpha \approx 2.9$, $\md \approx 4.03\times 10^{27}\g$,
            $\sigmap\approx 4.90\times 10^4\cm^{-2}$, $\chi^2/N\approx 1.81$).
        Lower panel (b): dotted line (model no.\,3) -- 
            best-fit $P=0.80$ ($\Pice\approx 0.46$) model
            ($\alpha \approx 3.0$, $\md \approx 5.73\times 10^{27}\g$,
            $\sigmap\approx 7.55\times 10^4\cm^{-2}$, $\chi^2/N\approx 6.08$);
            solid line (model no.\,4) -- 
            best-fit $P=0.60$ ($\Pice\approx 0$) model
            ($\alpha \approx 3.2$, $\md \approx 4.34\times 10^{27}\g$,
            $\sigmap\approx 1.31\times 10^5\cm^{-2}$, $\chi^2/N\approx 7.60$).
         }
\end{figure}

We illustrate in Figure \ref{fig:cold} the best
fits obtained for dust with a porosity of $P=0.95,0.90,0.80,0.60$.
Model parameters and results are tabulated in Table \ref{tab:para}. 
It is seen that the best fit is provided by dust with 
$P\simeq 0.90$ (with a total mass of $\approx 0.67\,m_\oplus$; model no.\,2) 
-- dust more porous than this is somewhat too hot
so that its emission is deficient in the submillimeter 
wavelength range (see Figure \ref{fig:cold}a);
on the other hand, dust more compact than this is 
a bit too cold so that it produces too much emission 
in this wavelength range (see Figure \ref{fig:cold}b).

\subsection{``Hot-Nebula'' Dust Model\label{sec:hot}}
Now we consider another extreme case: dust in the inner disk
regions has been so heavily processed that all amorphous silicate 
dust has been annealed ($T\simgt 800\K$) and all carbonaceous dust 
has been oxidized ($T\simgt 1100\K$) by reacting with OH. 
The annealed silicate dust is then transported to
(or the silicate vapour recondenses in crystalline form in)
the cold outer regions of the disk where these grains grow
into fluffy aggregates and are ultimately built into
planetesimals/cometesimals. 
Apparently, only a fraction of the dust in the disk is accreted to 
the inner warm regions and, hence, only a fraction of the silicate 
dust is crystalline. But since we are considering
an ``extreme'' case, in this section we model the ``hot-nebula'' 
dust as porous aggregates of {\it pure} crystalline silicate dust 
(plus ices in cold regions where dust reaches a temperature
of $\sim 110-120\K$). We do not know how much ice would 
recondense on the silicate core seeds in the fluffy aggregate,
but the assumption of full condensation of all condensible 
volatile elements (C, O, N) as ices (see \S\ref{sec:composition}) 
seems to be at the highest end. As a complement to \S\ref{sec:cold} 
where full condensation is assumed, we take $\vice/\vsil =1$,
a plausible value for dense molecular clouds.

\begin{figure}[h]
\begin{center}
\epsfig{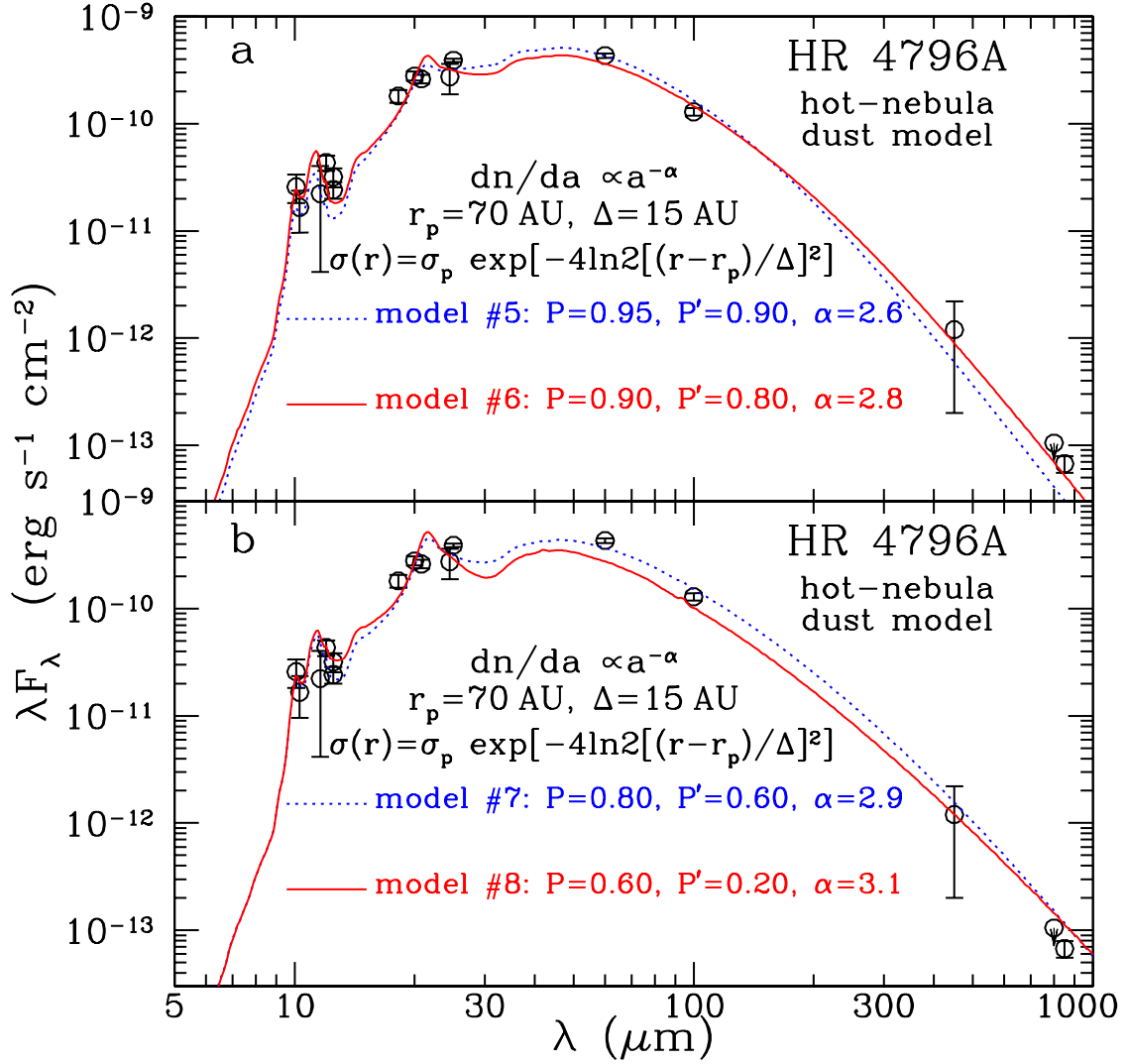}
\end{center}\vspace*{-1em}
\caption{
        \label{fig:hot}
        \footnotesize        
        Same as Figure \ref{fig:cold} but for the ``hot-nebula'' model 
        (porous aggregates of crystalline silicate dust plus ices
        in cold regions). 
        Upper panel (a): dotted line (model no.\,5) -- 
            best-fit $P=0.95$ ($\Pice\approx 0.90$) model
            ($\alpha \approx 2.6$, $\md \approx 6.02\times 10^{27}\g$,
            $\sigmap\approx 8.54\times 10^3\cm^{-2}$, $\chi^2/N\approx 5.29$);
            solid line (model no.\,6) -- 
            best-fit $P=0.90$ ($\Pice\approx 0.80$) model
            ($\alpha \approx 2.8$, $\md \approx 7.48\times 10^{27}\g$,
            $\sigmap\approx 2.55\times 10^4\cm^{-2}$, $\chi^2/N\approx 4.48$).
        Lower panel (b): dotted line (model no.\,7) -- 
            best-fit $P=0.80$ ($\Pice\approx 0.60$) model
            ($\alpha \approx 2.9$, $\md \approx 1.16\times 10^{28}\g$,
            $\sigmap\approx 4.30\times 10^4\cm^{-2}$, $\chi^2/N\approx 5.24$);
            solid line (model no.\,8) -- 
            best-fit $P=0.60$ ($\Pice\approx 0.20$) model
            ($\alpha \approx 3.1$, $\md \approx 9.25\times 10^{27}\g$,
            $\sigmap\approx 8.04\times 10^4\cm^{-2}$, $\chi^2/N\approx 9.94$).
        }
\end{figure}

Again, model IR spectra are calculated for a wide range of 
dust porosities $P$ and a wide range of dust power-indices $\alpha$.
Except the sharp features at 11.3$\mum$ and 23$\mum$, 
the ``hot-nebula'' model spectra are very similar to those of 
the ``cold-coagulation'' model. 
For illustration, we show in Figure \ref{fig:hot} the best-fit
model spectra provided by dust with $P=0.95,0.90,0.80,0.60$.
Model parameters and results are also tabulated in Table \ref{tab:para}.
Similar to the cold-coagulation model, the best fit is given by 
dust with $P\simeq 0.90$ (with a total mass of 
$\approx 1.25\,m_\oplus$; model no.\,6).

\begin{table}[h,t]
{\tiny
\caption[]{Models for the $\hra$ dust disk IR emission.\tablenotemark{a}\label{tab:para}}
\begin{tabular}{llllllllllllllll}
\hline \hline
	model 
        &dust 
        &spatial
	&$P$
        &$\Pice$

	&$\Delta$
        &$\beta$

      	& $\alpha$

      	& $\langle a\rangle$\tablenotemark{c}
	& $\langle a^2\rangle$\tablenotemark{c}
	& $\langle a^3\rangle$\tablenotemark{c}
        & $m_{\rm d}$
	& $\sigmap$
        &$\tau_{\rm p}^{V}$\tablenotemark{d}
      	& $\chi^2/N$ 
        & note
	\\
        no.
        &type
        &distrib.\tablenotemark{b}
      	&
        & 
        &(AU) 
        &
        & 

      	&($\mu$m)
	&(${\rm \mu m}^2$)
	&($10^{-8}\cm^3$)
        &($10^{27}\g$)
	&($10^{4}\cm^{-2}$)
	&  
	&\\ 
\hline
1	& cold-coag. &Gaussian& 0.95 & 0.87 
        & 15 & ... 
        & 2.8 
        & 2.25 & 47.8 & 9.46 
        & 2.67 & 2.97 
        & 0.041 & 2.85 & \\ 

2	& cold-coag. &Gaussian & 0.90 & 0.73 
        & 15 & ... 
        & 2.9 
        & 2.11 & 28.7 & 4.34 
        & 4.03 & 4.90 
        & 0.042 & 1.81 & preferred\\ 

3	& cold-coag. & Gaussian & 0.80 & 0.46 
        & 15 & ... 
        & 3.0 
        & 2.00 & 18.4 & 2.00 
        & 5.73 & 7.55 
        & 0.043 & 6.08 & \\ 

4	& cold-coag. & Gaussian & 0.60 & 0 
        & 15 & ... 
        & 3.2 
        & 1.83 & 9.26 & 0.44 
        & 4.34 & 13.1 
        & 0.038 & 7.60 & \\ 

5	& hot-nebu. & Gaussian & 0.95 & 0.90 
        & 15 & ... 
        & 2.6 
        & 2.66 & 155 & 45.5 
        & 6.02 & 0.85 
        & 0.040& 5.29 & \\ 

6	& hot-nebu. & Gaussian & 0.90 & 0.80 
        & 15 & ... 
        & 2.8 
        & 2.25 & 47.8 & 9.46 
        & 7.48 & 2.55 
        & 0.036& 4.48 & preferred\\ 

7	& hot-nebu. & Gaussian & 0.80 & 0.60 
        & 15 & ... 
        & 2.9 
        & 2.11 & 28.7 & 4.34 
        & 11.6 & 4.30 
        & 0.037&5.24 & \\ 

8	& hot-nebu. & Gaussian & 0.60 & 0.20 
        & 15 & ... 
        & 3.1 
        & 1.91 & 12.6 & 0.93 
        & 9.25 & 8.04 
        & 0.031&9.94 & \\ 

9	& cold-coag. & power-law & 0.90 & 0.73 
        & ... & -2.3 
        & 3.2 
        & 1.83 & 9.26 & 0.44 
        & 8.43 & 0.42 
        & 0.0011& 2.32 & \\ 

10	& cold-coag. & power-law & 0.90 & 0.73 
        & ... & -2.5 
        & 3.0 
        & 2.00 & 18.4 & 2.00 
        & 5.81 & 0.98 
        & 0.0054&3.28 & $r\in [40,130{\rm AU}]$\\ 

11	& hot-nebu. & power-law & 0.90 & 0.80 
        & ... & -1.1 
        & 2.9 
        & 2.11 & 28.7 & 4.34 
        & 25.5 & 0.26 
        & 0.0022& 9.71 & \\ 

12	& hot-nebu. & power-law & 0.90 & 0.80 
        & ... & 5.2 
        & 2.6 
        & 2.66 & 155 & 45.5 
        & 6.29 & 0.056 
        & 0.0026&2.02 & $r\in [40,130{\rm AU}]$ \\ 

13	& cold-coag. & Telesco & 0.90 & 0.73 
        & 21 & ... 
        & 2.9 
        & 2.11 & 28.7 & 4.34 
        & 5.41 & 2.34 
        & 0.020&3.60 & \\ 

14	& hot-nebu. & Telesco & 0.90 & 0.80 
        & 21 & ... 
        & 2.8 
        & 2.25 & 47.8 & 9.46 
        & 10.0 & 1.22 
        & 0.017&4.89 & \\ 

15	& cold-coag. & Gaussian & 0.90 & 0.73 
        & 15 & ... 
        & 2.8 
        & 0.22 & 0.81 & 0.15 
        & 5.49 & 193 
        & 0.045&3.44 & $\amin=0.1\mum$ \\ 

16	& cold-coag. & Gaussian & 0.90 & 0.73 
        & 15 & ... 
        & 3.1 
        & 19.1 & 1048 & 117 
        & 2.98 & 0.13 
        & 0.042&5.88 & $\amin=10\mum$ \\ 
\hline
\end{tabular}
\tablenotetext{a}{Unless otherwise stated, all models assume $\amin=1\mum$,
                  $\amax=1\cm$, $\rp=70\AU$, and $r\in [0.15,250\AU]$.}
\tablenotetext{b}{Dust spatial distribution --
                  ``Gaussian'': $\sigmar=\sigmap 
                  \exp[-4\ln2\{(r-\rp)/\Delta\}^2]$;
                  ``power-law'': $\sigmar = \sigmap 
                  \left(r/r_{\rm p}\right)^{-\beta}$;
                  ``Telesco'': $\sigmar= \sigmap 
                  \exp\left[-\left(\Delta/{\rm AU}\right) 
                  \ln^2\left(r/\rp\right)\right]$
                  for $r\in [0.15,70\AU]$;
                  $\sigmar=\sigmap/
                  \left\{1+\left[\left(r-\rp\right)/\Delta\right]^2\right\}$
                  for $r\in [70,105\AU]$;
                  $\sigmar=\sigmap \left[-0.01 \left(r/{\rm AU}\right) 
                  + 1.32\right]$ for $r\in [105,130\AU]$.}
\tablenotetext{c}{$\langle a^{\gamma}\rangle = \int_{\amin}^{\amax} da\
                   a^{\gamma}\ dn(a)/da/\int_{\amin}^{\amax} da\ dn(a)/da$.
                   The dust surface mass density can be written as
                   $\sigma(r;m)=\langle m\rangle\,\sigmar$ 
                   with the mean dust mass 
                   $\langle m \rangle =(4\pi/3)\langle a^3\rangle
                   \langle\rho\rangle$ 
                   where $\langle\rho\rangle \approx 2.5\,(1-P)\g\cm^{-3}$,
                   $\langle\rho^\prime\rangle \approx 1.7\,(1-\Pice)\g\cm^{-3}$
                   for the ``cold-coagulation'' model;
                   and $\langle\rho\rangle \approx 3.5\,(1-P)\g\cm^{-3}$,
                   $\langle\rho^\prime\rangle \approx 2.4\,(1-\Pice)\g\cm^{-3}$
                   for the ``hot-nebula'' model 
                   (see \S\ref{sec:porosity}).}
\tablenotetext{d}{Vertical optical depth at $\lambda=0.55\mum$
                  and $r=\rp$. The reason why $\tau_{\rm p}^V$ is much
                  smaller for models no.\,9--11 is that the bulk of the
                  dust is piled at the edge of the disk which is unphysical 
                  (see text [\S\ref{sec:pwl}]).}
}
\end{table}

\section{Discussion\label{sec:discussion}}
It is seen in \S\ref{sec:model} that, using a simple power-law 
dust size distribution ($dn/da\propto a^{-\alpha}$ 
with $\alpha \approx 2.8-2.9$, $\langle a\rangle \approx 2\mum$)
and a Gaussian-type dust spatial distribution with a peak
at $r_{\rm p}=70\AU$ and a FWHM $\Delta=15\AU$ as inferred 
from the near-IR imaging observation of scattered starlight 
(Schneider et al.\ 1999; Kenyon et al.\ 1999), 
both the ``cold-coagulation'' dust model (no.\,2) and 
the ``hot-nebula'' dust model (no.\,6) are successful in reproducing 
the observed SED, provided that the dust in the $\hra$ disk
is very fluffy, with a porosity $P\approx 0.90$. 
We note that, since we use this dust spatial distribution,
it is expected that our models are also able to reproduce 
the imaging observations.

Both the ``cold-coagulation'' model and the ``hot-nebula'' model
predict a vertical optical depth at visible wavelengths of 
$\tau^V\approx 0.04$ at $r=\rp$ (see Table \ref{tab:para}).
This justifies the optically-thin treatment employed 
in the entire paper.
This is because that in these models, with a power-law index for
the size distribution of $\alpha \in [2.6,3.2]$ 
(see Table \ref{tab:para}), most of the mass of 
the particles is in the largest grains.
However, the opacity per gram dust 
($\kappa_{\rm abs}\equiv \left[3/4\rho\right] Q_{\rm abs}/a$
where $Q_{\rm abs}$ is the absorption efficiency) 
for these macroscopic grains
($Q_{\rm abs}\approx 1$ at visible wavelengths)
is much smaller than that of micron-sized dust:
$\kappa_{\rm abs} \propto 1/a$.
We note that although the maximum dust size $\amax$
is not well constrained, the predicted IR emission spectrum
and the optical depth are not very sensitive to the precise
value of $\amax$ since grains larger than $\sim 100\mum$ 
have $Q_{\rm abs} \approx 1$ and emit like black-bodies.  

\subsection{Robustness\label{sec:robust}}
The extent to which dust in protostellar disks has been processed 
prior to incorporation into planetesimals and comets is not known. 
But it seems plausible that a fraction of precometary materials has
been processed in protostellar nebulae, based on the observationally
presence of crystalline silicate dust in comets (see Wooden 2002 for
a summary of the evidence). While at most $\sim 5\%$ of the silicate 
material is crystalline in the interstellar medium (Li \& Draine 2001a),
in some comets the fraction of crystalline silicate dust is as high 
as $\sim 30\%$. Accretion heating in protoplanetary disks is
a probable source for the formation site of the crystalline silicate
-- although $^{26}$Al heating of comets and consequent release of heat
in the amorphous-to-crystalline ice transition cannot be ruled out.
The warm inner zone of the accretion disk is particularly appealing 
as a site for the annealing of amorphous silicate. As a result of 
diffusional mixing induced by turbulence in the optically thick 
disk (Gail 2001), or outflows driven by reconnecting magnetic field 
lines (the ``X-winds'' model of Shu, Shang, \& Lee [1996]), 
a fraction of the annealed silicate dust arrives at the cooler outer 
part of the disk to be incorporated in icy ``cometary'' bodies.

Therefore, we expect that the dust generated by collisions
among cometary bodies (see \S\ref{sec:rppr}),
for either origin of the crystalline silicate material, 
should be intermediate between the ``cold-coagulation'' dust 
and the ``hot-nebula'' dust.
The mixing ratio of these two dust types is unclear. 
In principle, the 10$\mum$ silicate emission feature should
allow us to infer the thermal history of silicate dust through
the absence or presence (and the mass fraction) of crystalline
silicate dust. However, the only available 8--13$\mum$ spectrum
shows very weak thermal emission in the silicate feature;
its quality is inadequate to discern the detailed spectral 
features expected from crystalline silicate dust
(Sitko, Lynch, \& Russell 2000). But in any case, the fact that
the two extreme models are able to provide close fits to
the observed SED, any linear combinations of these two dust types 
(i.e., any degree of processing/modification) are expected to 
be able to reproduce the observed SED as well.
Therefore, our models are robust.

\subsection{Uniqueness\label{sec:pwl}}
In modelling SED, dust sizes and dust spatial distribution
are coupled. Since the dust spatial distribution we adopted
for the $\hra$ disk is well constrained by near-IR images of 
scattered light (Schneider et al.\ 1999), the dust properties 
are more or less uniquely determined: the dust must be very
fluffy (with $P\simeq 0.9$); the dust power-law size distribution
has an index $\alpha \approx 2.8-2.9$ with a mean dust size of
$\sim 2\mum$ (see Table \ref{tab:para}); the dust composition 
should be somewhere between the ``cold-coagulation'' dust 
and the ``hot-nebula'' dust although its detailed composition
(e.g. the crystallinity of the silicate dust) is yet unknown
(see \S\ref{sec:robust}).

\begin{figure}[h]
\begin{center}
\epsfig{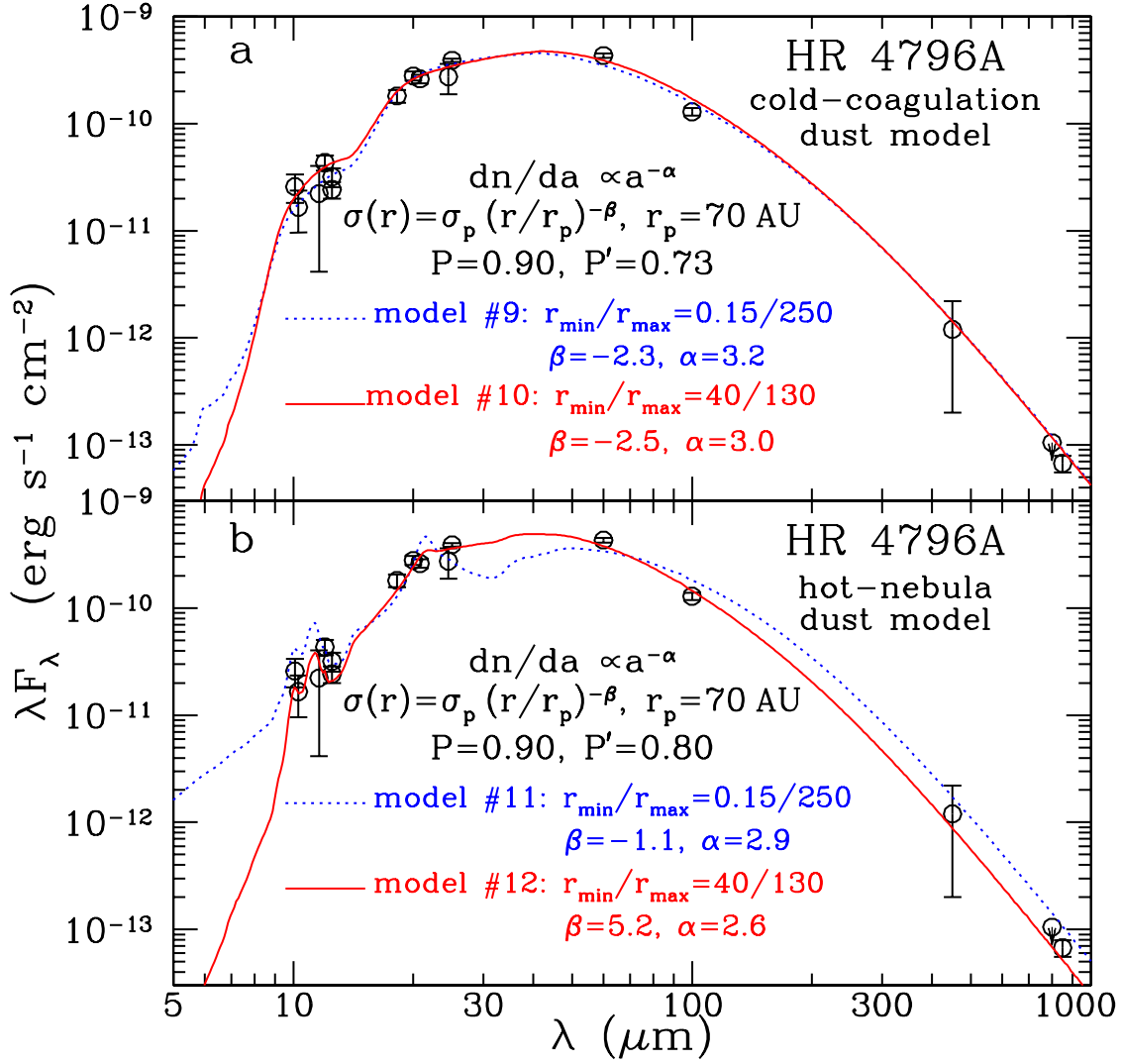}
\end{center}\vspace*{-1em}
\caption{
        \label{fig:pwl}
        \footnotesize
        Comparison of the $\hra$ observational SED 
        to theoretical IR emission spectra
        calculated from (a) the ``cold-coagulation'' model
        and (b) the ``hot-nebula'' model.
        The dust spatial distribution is assumed to 
        be a power-law $\sigmar = \sigmap \left(r/r_{\rm p}\right)^{-\beta}$
        for either $0.15\simlt r \simlt 250\AU$ (dotted lines)
        or $40\simlt r \simlt 130\AU$ (solid lines).
        We assume a power law dust size distribution 
        $dn(a)/da \propto a^{-\alpha}$ ($a\in [1\mum,1\cm]$).
        Upper panel (a): dotted line (model no.\,9) -- 
            best-fit $P=0.90$ ($\Pice\approx 0.73$)
            ``cold-coagulation'' dust model for $r\in [0.15,250\AU]$
            ($\beta\approx -2.3$, $\alpha \approx 3.2$, 
            $\md \approx 8.43\times 10^{27}\g$, 
            $\sigmap\approx 4.17\times 10^3\cm^{-2}$, 
            $\chi^2/N\approx 2.32$);
        solid line (no.\,10) -- best-fit $P=0.90$ ($\Pice\approx 0.73$)
            ``cold-coagulation'' dust model for $r\in [40,130\AU]$ 
            ($\beta\approx -2.5$, $\alpha \approx 3.0$, 
            $\md \approx 5.81\times 10^{27}\g$, 
            $\sigmap\approx 9.75\times 10^3\cm^{-2}$,  
            $\chi^2/N\approx 3.28$).
        Lower panel (b): dotted line (model no.\,11) -- 
            best-fit $P=0.90$ ($\Pice\approx 0.80$)
            ``hot-nebula'' dust model for $r\in [0.15,250\AU]$
            ($\beta\approx -1.1$, $\alpha \approx 2.9$, 
            $\md \approx 2.55\times 10^{28}\g$,
            $\sigmap\approx 2.58\times 10^3\cm^{-2}$,  
            $\chi^2/N\approx 9.71$);
        solid line (model no.\,12) -- 
            best-fit $P=0.90$ ($\Pice\approx 0.80$)
            ``hot-nebula'' dust model for $r\in [40,130\AU]$
            ($\beta\approx 5.2$, $\alpha \approx 2.6$, 
            $\md \approx 6.29\times 10^{27}\g$,
            $\sigmap\approx 5.55\times 10^2\cm^{-2}$,   
            $\chi^2/N\approx 2.02$).
        }
\end{figure}

The dust spatial distribution in the $\hra$ disk was often
modelled as a single power-law (Jayawardhana et al.\ 1998;
Koerner et al.\ 1998; Wyatt et al.\ 1999) or two power-law
components (Augereau et al.\ 1999). For the purpose of comparison,
we have also modelled the $\hra$ SED assuming a single power-law 
dust distribution $dn/dr \propto r^{-\beta}$ in terms of 
the $P=0.90$ ($\Pice\approx 0.73$) ``cold-coagulation'' dust model 
and the $P=0.90$ ($\Pice\approx 0.80$) ``hot-nebula'' dust model.\footnote{%
  As in \S\ref{sec:method}, the dust surface density distribution
  is written as $\sigmar = \sigmap \left(r/r_{\rm p}\right)^{-\beta}$.
  }   
We consider two disk extents: 
(1) $\rin=0.15 \simlt r \simlt \rout=250\AU$ (see \S\ref{sec:method}); 
and (2) $\rin=40 \simlt r \simlt \rout=130\AU$.
The latter takes into account the existence of an inner hole
with a radius of $\sim 40-60\AU$ which was first noticed by 
Jura et al.\ (1995) from an analysis of the IRAS ({\it Infrared 
Astronomical Satellite}) and ground-based photometry.
The inner hole was later confirmed by mid-IR imaging 
of dust thermal emission carried out by three independent groups 
(Jayawardhana et al.\ 1998; Koerner et al.\ 1998; Telesco et al.\ 2000).
These mid-IR imaging observations also implied an outer radius of 
$\sim 110-130\AU$ for the $\hra$ disk.
As can be seen in Figure \ref{fig:pwl}, models with a single
power-law dust spatial distribution are also successful in
reproducing the observed SED. However, except the 
$40 \simlt r \simlt 130\AU$ ``hot-nebula'' model (no.\,12),
all models (no.\,9 -- no.\,11) require an increasing 
accumulation of dust toward the outer edge (i.e. $\beta < 0$). 
This is neither physical nor consistent with 
the imaging observations of scattered light 
(Schneider et al.\ 1999) and dust thermal emission 
(Jayawardhana et al.\ 1998; Koerner et al.\ 1998; Telesco et al.\ 2000).

It is expected that models with two power-law dust spatial 
distributions are also able to reproduce the observed SED.
But the constraints placed by the near-IR and mid-IR imaging 
observations would imply that the two-segment power-law 
distribution should not deviate much from the Gaussian function.

\begin{figure}[h]
\begin{center}
\epsfig{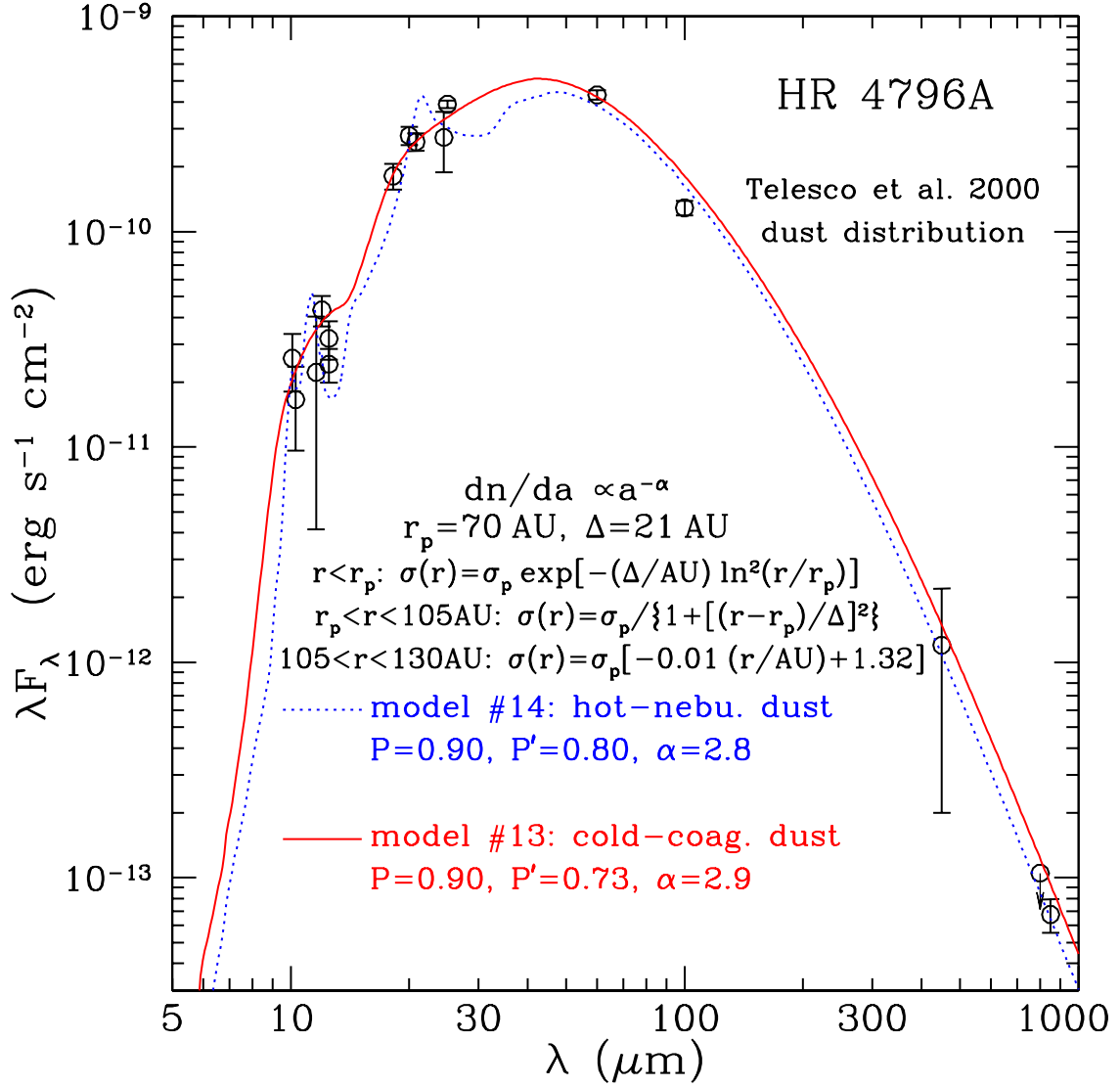}
\end{center}\vspace*{-1em}
\caption{
        \label{fig:telesco}
        \footnotesize
        Comparison of the $\hra$ observational SED 
        to theoretical IR emission spectra
        calculated from the ``cold-coagulation'' model
        (solid line [no.\,14]; $P=0.90$, $\Pice\approx 0.73$,
         $\alpha \approx 2.9$, 
         $\md \approx 5.41\times 10^{27}\g$,
         $\sigmap\approx 2.34\times 10^4\cm^{-2}$,  
         $\chi^2/N\approx 3.60$)
        and the ``hot-nebula'' model (dotted line [no.\,13];
         $P=0.90$, $\Pice\approx 0.80$,
         $\alpha \approx 2.8$, 
         $\md \approx 1.00\times 10^{28}\g$,
         $\sigmap\approx 1.22\times 10^4\cm^{-2}$,  
         $\chi^2/N\approx 4.89$).
        The dust spatial distribution is taken to 
        be that of Telesco et al.\ (2000).
        }
\end{figure}

Assuming spherical silicate dust of a diameter of 2.5$\mum$
(estimated from the 10.8$\mum$ and 18.2$\mum$ emission),
Telesco et al.\ (2000) derived the dust spatial distribution
from the 18.2$\mum$ brightness distribution (see their Figure 5).
We approximate their result as: for $r < \rp=70\AU$, 
$\sigma \approx \sigmap \exp\left[-\left(\Delta/{\rm AU}\right) 
\ln^2\left(r/\rp\right)\right]$ ($\Delta \approx 21\AU$); 
for $70\AU < r < 105\AU$, $\sigma \approx \sigmap/
\left\{1+\left[\left(r-\rp\right)/\Delta\right]^2\right\}$;
for $105\AU < r < 130\AU$, 
$\sigma \approx \sigmap \left[-0.01 \left(r/{\rm AU}\right) + 1.32\right]$.
In Figure \ref{fig:telesco} we show that the best-fitting 
``cold-coagulation'' (no.\,13) and ``hot-nebula'' models (no.\,14)
with this dust distribution also provide close fits to the observed SED.

The overall shape of the Telesco et al.\ (2000) distribution 
resembles that of the Gaussian distribution, except that the 
former is relatively broader and has a flatter wing at $r>70\AU$.
Since the Gaussian distribution was derived from scattered
light images which reflect dust of all sizes 
while the Telesco et al.\ (2000) distribution was derived 
only from the 10.8$\mum$ and 18.2$\mum$ mid-IR-emitting
dust, we prefer the Gaussian distribution although the
real dust distribution may not be an exact Gaussian function.

\begin{figure}[h]
\begin{center}
\epsfig{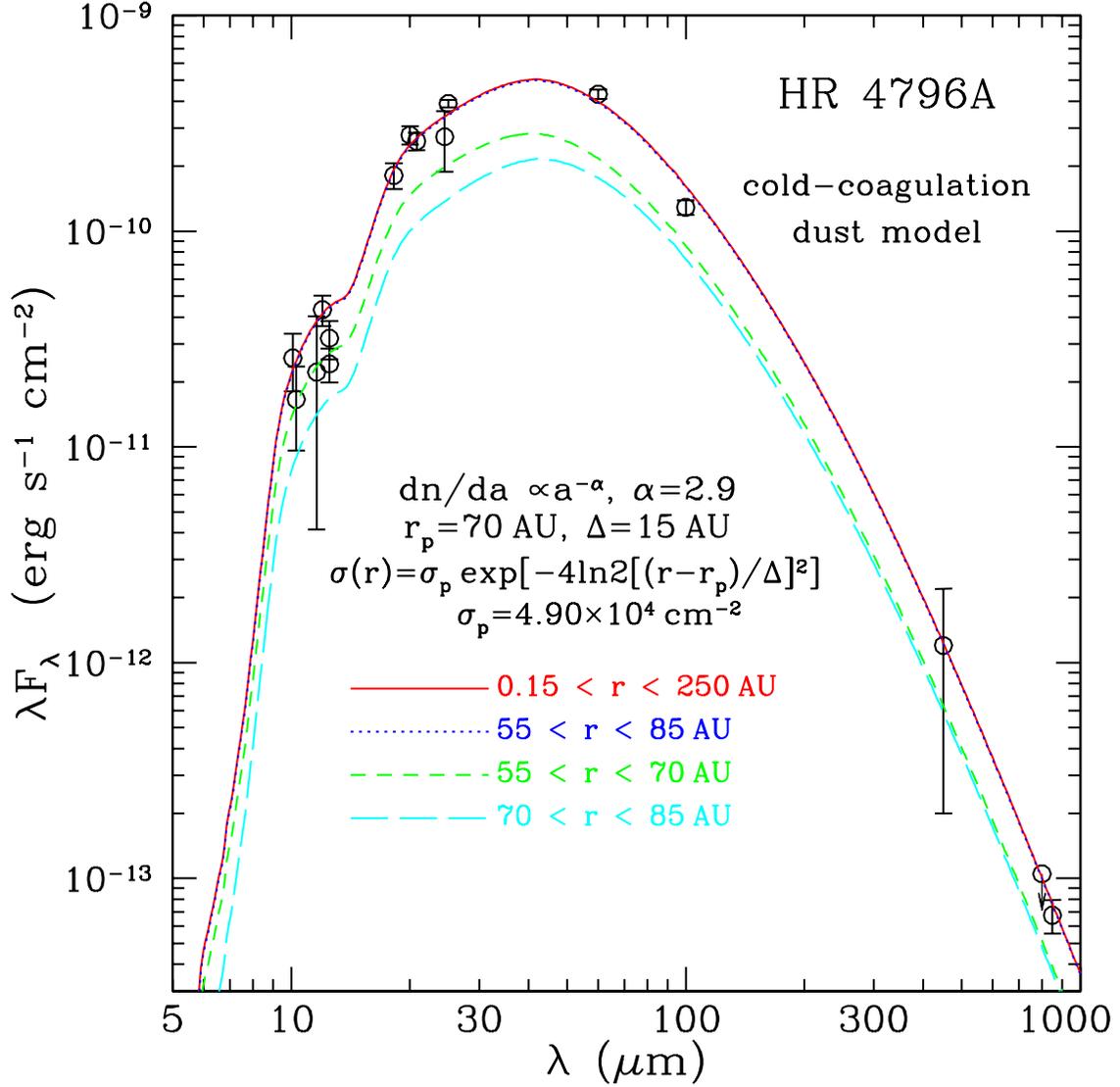}
\end{center}\vspace*{-1em}
\caption{
        \label{fig:diffrad}
        \footnotesize
        IR emission from dust in the ring regions
        of $\rin=0.15\AU < r < \rout=250\AU$ (solid line; 
        same as the solid line in Figure \ref{fig:cold}a),
        $(\rp-\Delta) < r < (\rp+\Delta)$ (dotted line),
        $(\rp-\Delta) < r < \rp$ (dashed line),
        and $\rp < r < (\rp+\Delta)$ (long-dashed line)
        calculated from the best-fitting ``cold-coagulation'' 
        dust model (no.\,2; $P=0.90$, $\Pice\approx 0.73$, 
        $\alpha\approx 2.9$) with a Gaussian spatial distribution 
        ($\rp=70\AU$, $\Delta=15\AU$).
        The $(\rp-\Delta) < r < (\rp+\Delta)$ spectrum (dotted line)
        is indistinguishable from the $0.15\AU < r < 250\AU$ 
        spectrum (solid line).
        }
\end{figure}

To illustrate the disk regional contribution,
we show in Figure \ref{fig:diffrad} the IR emission
from dust in the regions of $(\rp-\Delta) < r < \rp$,
$\rp < r < (\rp+\Delta)$, and $(\rp-\Delta) < r < (\rp+\Delta)$
as well as $0.15\AU < r < 250\AU$ calculated from the
best-fitting ``cold-coagulation'' dust model
(no.\,2; $P=0.90$, $\Pice\approx 0.73$, $\alpha\approx 2.9$)
with a Gaussian spatial distribution
($\rp=70\AU$, $\Delta=15\AU$).
It is seen that the IR emission is exclusively contributed
by dust confined in the ring region of 
$(\rp-\Delta) < r < (\rp+\Delta)$.

\begin{figure}[h]
\begin{center}
\epsfig{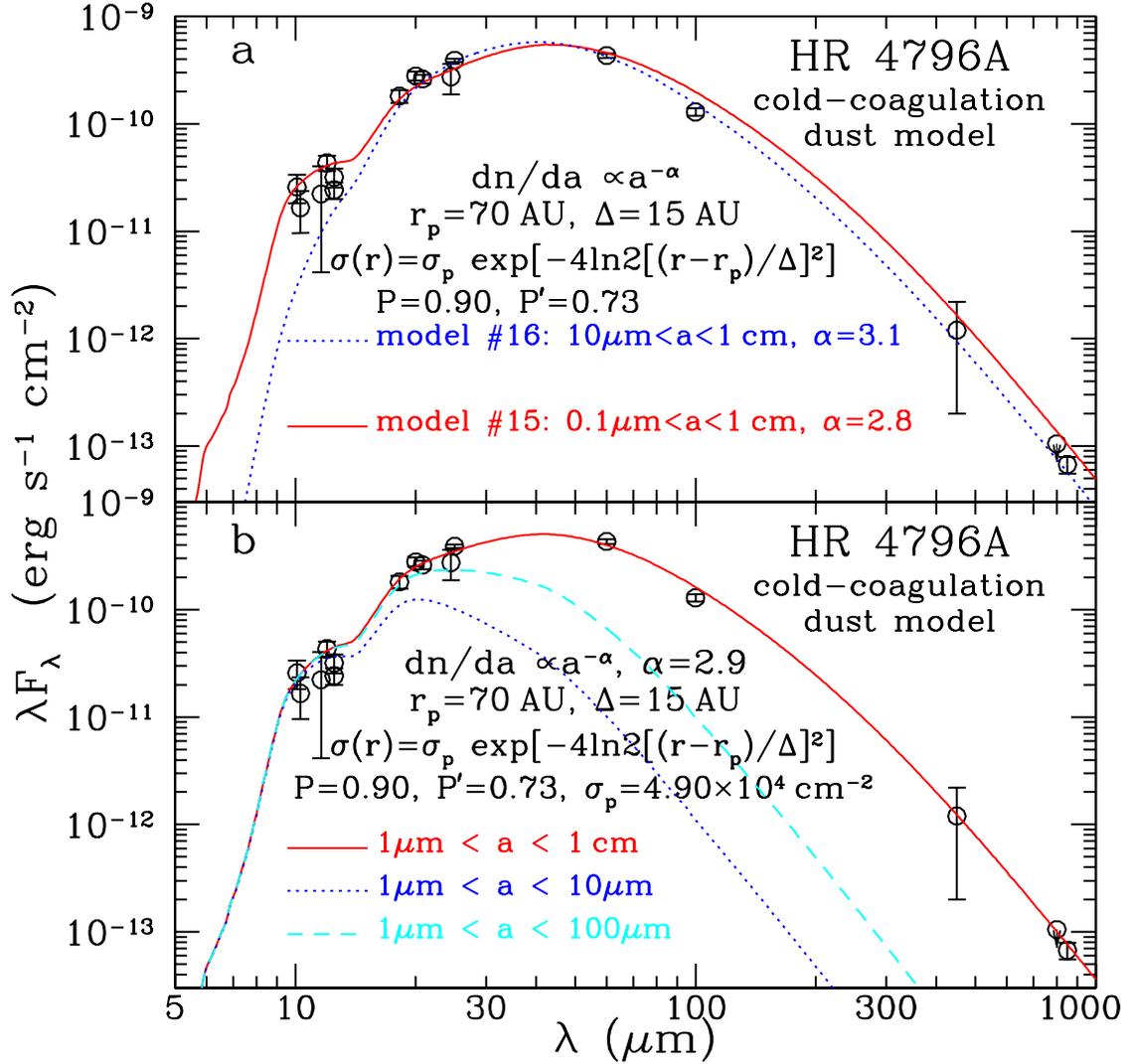}
\end{center}\vspace*{-1em}
\caption{
        \label{fig:diffsz}
        \footnotesize
        Upper panel (a): models with smaller $\amin$
        ($=0.1\mum$; solid line [no.\,15]; $\alpha \approx 2.8$, 
        $\md \approx 5.49\times 10^{27}\g$, 
        $\sigmap\approx 1.93\times 10^6\cm^{-2}$, 
        $\chi^2/N\approx 3.44$)
        and larger $\amin$ ($10\mum$; dotted line [no.\,16]; 
        $\alpha \approx 3.1$, 
        $\md \approx 2.98\times 10^{27}\g$, 
        $\sigmap\approx 1.35\times 10^3\cm^{-2}$, 
        $\chi^2/N\approx 5.88$). Other model parameters
        are the same as those of the best-fit ``cold-coagulation''
        model (no.\,2; $P=0.90$, $\Pice\approx 0.73$).
        Lower panel (b): model spectra calculated from the
        best-fit ``cold-coagulation'' model 
        (no.\,2; $P=0.90$, $\Pice\approx 0.73$, $\alpha \approx 3.1$)
        with $1\mum <a <10\mum$ (dotted line),
        $1\mum <a <100\mum$ (dashed line),
        $1\mum <a <1\cm$ (solid line; same as the solid line in
        Figure \ref{fig:cold}a).
        }
\end{figure}

So far, all models assume $\amin=1\mum$.
We now consider models with smaller or larger $\amin$ values.
In Figure \ref{fig:diffsz}a we show the best-fit spectra calculated
from models with $\amin=0.1\mum$ (model no.\,15)
and $\amin=10\mum$ (model no.\,16). It is seen that 
(1) models with $\amin=0.1\mum$\footnote{%
  This requires that the constituent individual particles
  should be smaller than the typical interstellar dust 
  ($a\sim 0.1\mum$).
  }
emit a bit too much at $\lambda \simgt 100\mum$; 
decreasing the dust porosity improves
the fit at $\lambda \simgt 100\mum$ but the fit to 
the $\lambda \simlt 10\mum$ part deteriorates;
(2) models with $\amin=10\mum$ emit too little at 
$\lambda \simlt 10\mum$; increasing the dust porosity
does not solve this problem. 

To illustrate the IR emission contributed by different
dust sizes, we show in Figure \ref{fig:diffsz}b the emission
spectra calculated from the best-fitting ``cold-coagulation'' model 
(no.\,2; $P=0.90$, $\Pice\approx 0.73$, $\alpha\approx 2.9$)
with $1\mum < a <10\mum$ (subject to radiative expulsion; 
see \S\ref{sec:rppr}), $1\mum < a <100\mum$ (subject to 
Poynting-Robertson inward spiralling drag; see \S\ref{sec:rppr}),
and $1\mum < a <1\cm$. 
It is seen that the $\lambda \simlt 20\mum$ mid-IR emission
is dominantly produced by grains smaller than 100$\mum$ in radius.
Since these grains will be removed from the disk during its lifetime
either through radiative expulsion or through Poynting-Robertson
drag, there must exist a source of replenishment (see \S\ref{sec:rppr}). 

\begin{figure}[h]
\begin{center}
\epsfig{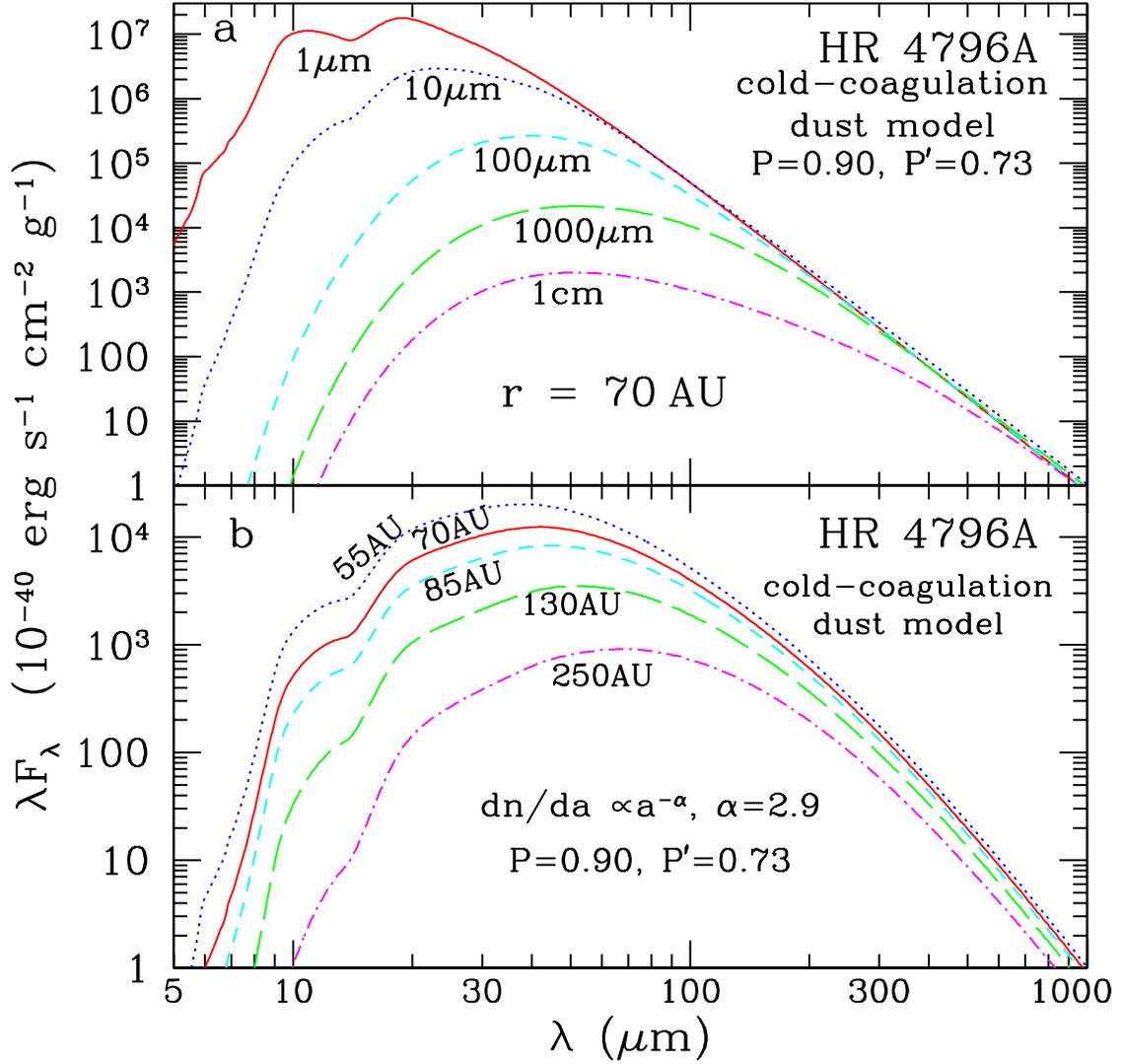}
\end{center}\vspace*{-1em}
\caption{
        \label{fig:radsz}
        \footnotesize
        Upper panel (a): IR emission per gram dust mass 
        (received at the Earth) calculated from
        the $P=0.90$ ($\Pice\approx 0.73$) ``cold-coagulation'' 
        dust at $r=70\AU$ with a single size of 
        $a=1\mum$ (solid line), $10\mum$ (dotted line), 
        $100\mum$ (dashed line), $1000\mum$ (long-dashed line),
        and $1\cm$ (dot-dashed line).
        Lower panel (b): IR emission per gram dust mass calculated 
        from the $P=0.90$ ($\Pice\approx 0.73$) ``cold-coagulation'' 
        dust at $r=55\AU$ (dotted line), $70\AU$ (solid line),
        $85\AU$ (dashed line), $130\AU$ (long-dashed line), 
        and $250\AU$ (dot-dashed line) with a distribution of sizes 
        ($dn/da \propto a^{-\alpha}$, $\alpha =2.9$,
        $a\in [1\mum, 1\cm]$).
        }
\end{figure}

Finally, we show in Figure \ref{fig:radsz}a the IR emission
per unit dust mass received at the Earth produced by the $P=0.90$ 
($\Pice\approx 0.73$) ``cold-coagulation'' dust at $r=70\AU$
with a single size of $a=1\mum, 10\mum, 100\mum, 1000\mum, 1\cm$ 
(at $r=70\AU$, all grains are colder than 110$\K$ and have 
a porosity of $\Pice=0.73$ [their constituent particles are 
coated by a layer of ice mantles] except the $a=1\mum$ dust 
which attains an equilibrium temperature of $\approx 154\K$ 
and thus has no ice mantles and has a porosity of $P=0.90$). 
In Figure \ref{fig:radsz}b we show the IR spectra emitted by 
a gram of the $P=0.90$ ($\Pice\approx 0.73$) ``cold-coagulation'' 
dust at $r=55\AU, 70\AU, 85\AU, 130\AU, 250\AU$ with a distribution 
of sizes ($dn/da \propto a^{-\alpha}$, $\alpha =2.9$,
$a\in [1\mum, 1\cm]$).

\subsection{Is There A ``Zodiacal Dust'' Component?\label{sec:zodi}} 
In modelling the $\hra$ SED, Koerner et al.\ (1998) and 
Augereau et al.\ (1999) argued that, in addition to the
dust responsible for the emission at $\lambda >20\mum$, 
a hot ``zodiacal dust'' component which is confined within 
$\sim 3-6\AU$ or $\sim 9-10\AU$ of the star is required to
account for the entire emission at $\lambda \simlt 12\mum$ and
$\simlt 5-10\%$ of the emission at $\lambda \sim 20\mum$.
However, it can be seen clearly from Figure \ref{fig:cold},
Figure \ref{fig:hot}, and even the unphysical single power-law
spatial distribution model of Figure \ref{fig:pwl} that, 
our models closely reproduce the observed SED over the entire 
wavelength range. There is thus no need to invoke a hot
``zodiacal'' dust component; indeed we may now ask how much
such dust can be tolerated based on our models.

\begin{figure}[h]
\begin{center}
\epsfig{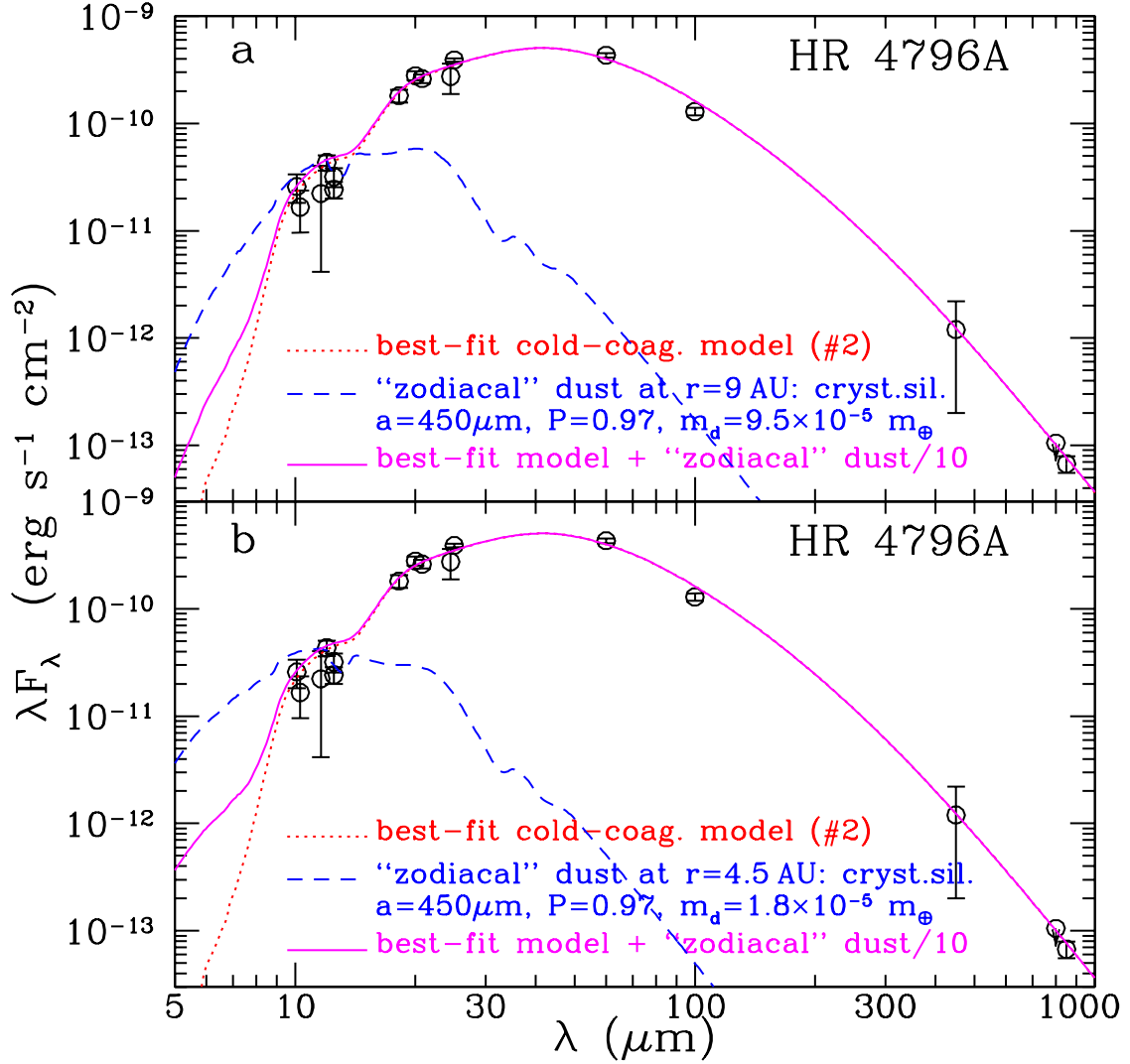}
\end{center}\vspace*{-1em}
\caption{
        \label{fig:zodi}
        \footnotesize
        Upper limits on the ``zodiacal'' dust component
        (a) at $r=9\AU$ ($m_{\rm d} \approx 9.5\times 10^{-6}\,m_\oplus$) 
        and (b) at $r=4.5\AU$ 
        ($m_{\rm d} \approx 1.8\times 10^{-6}\,m_\oplus$):
        dashed lines -- model spectra calculated from the ``zodiacal dust''
        which is taken to be porous crystalline silicate dust with
        $P=0.97$ and $a=450\mum$ (see Augereau et al.\ 1999) of (a)
        $m_{\rm d} \approx 9.5\times 10^{-5}\,m_\oplus$ at $r=9\AU$ 
        and (b) $m_{\rm d} \approx 1.8\times 10^{-5}\,m_\oplus$ 
        at $r=4.5\AU$, assuming the emission at $\lambda \simlt 12\mum$
        is entirely produced by the ``zodiacal'' dust;
        dotted lines -- the best-fit ``cold-coagulation'' model spectra 
        (no.\,2; $P=0.90$, $\Pice\approx 0.73$, $\alpha \approx 2.9$;
        same as the solid line in Figure \ref{fig:cold}a);
        solid lines -- the sum of 10\% of the ``zodiacal'' dust spectrum
        and the best-fit ``cold-coagulation'' model spectrum.                  
        }
\end{figure}

An upper limit on the ``zodiacal'' dust component can be obtained 
such that this abundance of hot dust, after added to the IR emission 
spectra of our best-fitting models (e.g., no.\,2), would not exceed the
observed emission at $\lambda \simlt 12\mum$.
Following Augereau et al.\ (1999), we assume that the ``zodiacal'' dust
consists of porous ($P\approx 0.97$) crystalline silicate grains
of $a=450\mum$. As shown in Figure \ref{fig:zodi}, if the observed 
emission at $\lambda \simlt 12\mum$ is entirely produced 
by the ``zodiacal'' dust, a total amount of 
$m_{\rm d} \approx 9.5\ (1.8)\times 10^{-5}\,m_\oplus$
is required if the ``zodiacal'' dust lies at $r=9\ (4.5)\AU$.
It is seen in Figure \ref{fig:zodi} that our best-fitting model (no.\,2)
allows at most 10\% of this mass to be in the ``zodiacal'' dust component.  

The reason why Koerner et al.\ (1999) required a population of
``zodiacal dust'' can be ascribed to the fact that they have not
considered a distribution of dust sizes; instead, they simply
approximated the dust emissivity $\epsilon_{\lambda}$ as a one-parameter 
function:  $\epsilon_{\lambda}=1$ for $\lambda < a_0$
and $\epsilon_{\lambda}=a_0/\lambda$ for $\lambda > a_0$
where $a_0$ is a parameter characteristics of dust size.
It would not be surprising if their model lacks
hot dust emitting at $\lambda \simlt 12\mum$
since their approximation is essentially for single-sized dust models. 

The reason why Augereau et al.\ (1999) needed a population of
``zodiacal dust'' may lie in the fact that they fixed the 
power-index of their dust size distribution to be $\alpha=3.5$.
Using a steeper size distribution 
(in comparison with $\alpha=2.8-2.9$ of our $P=0.90$ best-fitting 
models [no.\,2 and no.\,6; see Figures \ref{fig:cold},\ref{fig:hot}],
Augereau et al.\ (1999) had to raise $\amin$ to $\simgt 10\mum$ in 
order to reproduce the long wavelength emission. This led to a paucity 
of small grains which are hot enough to emit at $\lambda \simlt 12\mum$
(this can also be seen in Figure \ref{fig:diffsz}a [dotted line]).

Also arguing against attributing the $\lambda \simlt 12\mum$ emission
to a hot ``zodiacal dust'' component are the recent 10.8$\mum$ 
and 18.2$\mum$ mid-IR images of the $\hra$ disk that show that
the disk's size at 10$\mum$ is comparable to its size at 18$\mum$
(Telesco et al.\ 2000).
This implies that the 18$\mum$-emitting dust may also emit some, 
or all, of the 10$\mum$ radiation (Telesco et al.\ 2000).
This is in sharp contrast to the ``zodiacal dust'' assumption 
which suggested that the ``zodiacal dust'' component contributes 
to only $\simlt 5-10\%$ of the 20$\mum$ emission
(Koerner et al.\ 1998; Augereau et al.\ 1999).

The tidal effects by sweeping inner planets together with 
the Poynting-Robertson drag may be responsible for the absence 
of a warm ``zodiacal dust'' population as well as the presence 
of an inner disk hole ($r\simlt 50\AU$).

\subsection{Radiation Pressure and Poynting-Robertson Drag\label{sec:rppr}}
In addition to the gravitational attraction from the central star,
grains in the $\hra$ disk are subject to (1) radiative repulsion due to 
the momentum carried by stellar photons and (2) Poynting-Robertson drag 
which takes both energy and momentum from their orbits and causes them
to spiral toward the gravitational force center (Burns, Lamy, \& Soter 1979;
Backman \& Paresce 1993).

\begin{figure}[h]
\begin{center}
\epsfig{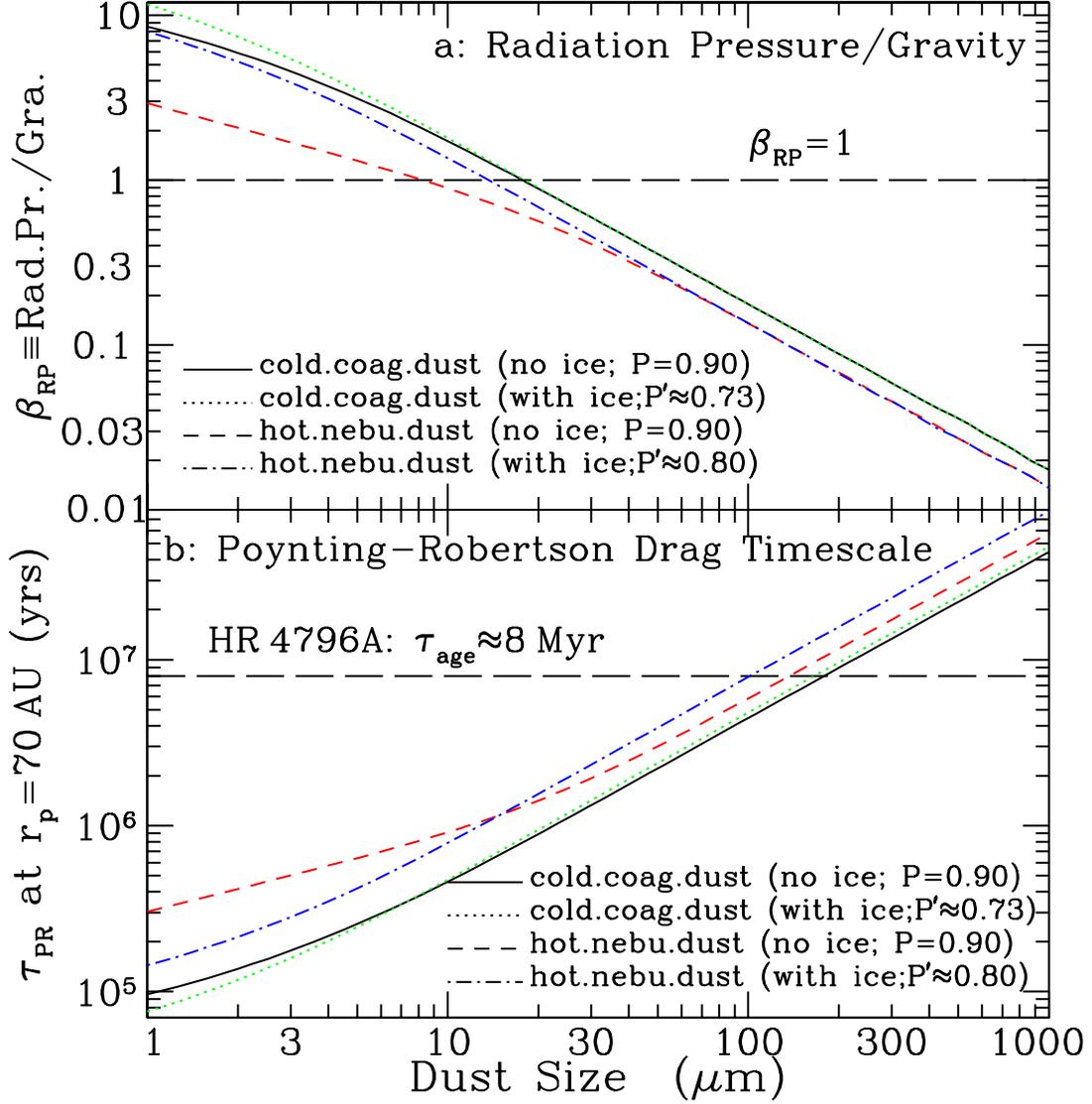}
\end{center}\vspace*{-1em}
\caption{
        \label{fig:rppr}
        \footnotesize
        Upper panel (a): the ratios of the radiative repulsion to 
        the gravitational attraction ($\beta_{\rm RP}$) for 
        the best-fitting ``cold-coagulation'' dust 
        (without ice $P=0.90$ [solid line] 
        or with ice $\Pice=0.73$ [dotted line])
        and the ``hot-nebula'' dust
        (without ice $P=0.90$ [dashed line] 
        or with ice $\Pice=0.80$ [dot-dashed line]).
        The long-dashed horizontal line plots $\beta_{\rm RP}=1$. 
        Lower panel (b): the orbit decay timescales $\tau_{\rm PR}$
        due to the Poynting-Robertson drag for 
        the best-fitting ``cold-coagulation'' dust
        (without ice $P=0.90$ [solid line] 
        or with ice $\Pice=0.73$ [dotted line])
        and the ``hot-nebula'' dust
        (without ice $P=0.90$ [dashed line] 
        or with ice $\Pice=0.80$ [dot-dashed line])
        at a radial distance of $r=70\AU$ from the central star
        (note $\tau_{\rm PR}\propto r^{2}$).
        The long-dashed horizontal line plots the $\hra$ age
        ($\approx 8\myr$).        
        }
\end{figure}

We have calculated $\beta_{\rm RP}$ -- the ratio of 
radiative pressure (RP) force to gravitational force 
for the best-fitting ``cold-coagulation'' dust 
(no.\,2; $P=0.90$, $\Pice\approx 0.73$) and 
``hot-nebula'' dust (no.\,6; $P=0.90$, $\Pice\approx 0.80$).
As shown in Figure \ref{fig:rppr}a, for grains smaller than 
$\sim 10-20\mum$, the radiation pressure overcomes the
gravitational attraction (i.e. $\beta_{\rm RP}\simgt 1$) 
and, therefore, these grains will be blown out from the $\hra$ disk.
The dust removal rate due to the radiation pressure expulsion
can be estimated by integrating
$\left(4\pi/3\right)\rho a^3/\tau_{\rm RP}(r)$ 
over the size range $1\mum<a<10\mum$ and over the entire disk.
Assuming the RP timescale is comparable to the local dynamical
timescale [$\tau_{\rm RP}(r)\approx 370\,\left(r/70\AU\right)^{3/2}\yr$],  
we estimate the RP dust mass loss rate to be 
$\approx 8.19\times 10^{-7}\,m_\oplus/{\rm yr}$ 
for the best-fitting ``cold-coagulation'' model (no.\,2) 
and $\approx 8.28\times 10^{-7}\,m_\oplus/{\rm yr}$ 
for the best-fitting ``hot-nebula'' model (no.\,6). 

We have also calculated the Poynting-Robertson (PR) drag timescales
($\tau_{\rm PR}$). In Figure \ref{fig:rppr}b we show that,
for grains smaller than $\sim 100-200\mum$ at a radial distance of
$r=70\AU$ from the central star, their lifetimes (at $r=70\AU$) 
due to the PR drag effect are shorter than the $\hra$ age and hence, 
for grains in the range of $10-20\mum \simlt a\simlt 100-200\mum$, 
although stable against radiation pressure ejection,
the Poynting-Robertson drag could remove them from the disk.
Since the best-fitting models have $\simlt 30-40\%$ of the total
surface areas in grains smaller than $100\mum$, these grains
must be efficiently replenished by cascade collisions of 
planetesimals and larger grains. 
By integrating the PR dust removal rate
$\left(4\pi/3\right)\rho a^3/\tau_{\rm PR}(a,r)$ 
over the whole size range and over the entire disk, 
we estimate the PR dust mass loss rate to be 
$\approx 9.22\times 10^{-9}\,m_\oplus/{\rm yr}$ 
for the best-fitting ``cold-coagulation'' model (no.\,2)
and $\approx 7.96\times 10^{-9}\,m_\oplus/{\rm yr}$ 
for the best-fitting ``hot-nebula'' model (no.\,6). 

Therefore, over the life span of $\hra$, 
roughly $6.7\,m_\oplus$ of dust is lost by radiation pressure 
and Poynting-Robertson drag, which is about 3 times smaller 
than the estimation of Augereau et al.\ (1999; $\approx 20\,m_\oplus$). 
This is mainly because Augereau et al.\ (1999) attributed 
the dominant dust removal process to grain-grain collisions.
We consider this unlikely since grain-grain collisions would
not remove the dust from the disk, instead, 
they just re-distribute the dust over different size bins 
through fragmentation or sticking.\footnote{%
  Since the grain-grain collision timescale $\tau_{\rm coll}
  \approx \left\{2\pi\langle a^2\rangle\sigmar\Omega(r)\right\}^{-1}
  \approx 670\yr$
  at $r=70\AU$ [$\Omega(r)\equiv \left(GM_{\star}/r^3\right)^{1/2}$ 
  is the Keplerian frequency; $M_\star$ ($\approx 2.5\,m_\odot$) is 
  the stellar mass; $G$ is the Gravitation constant] is comparable 
  to the RP timescale $\tau_{\rm RP}\approx 370\yr$ (at $r=70\AU$), 
  a stable population of $a<10\mum$ dust
  can also be obtained through a cascade production during 
  collisions between larger grains. 
  }

It is well established (see e.g. Spangler et al.\ 2001,
Zuckerman \& Becklin 1993) that the amount of IR excess 
around main sequence stars decreases with the star's age. 
There are at least two possible explanations for this result:
(1) the particles are lost by the Poynting-Robertson drag effect;
(2) the particles have coalesced into larger bodies which are 
undetectable as the opacity per gram decreases 
($\kappa_{\rm abs} \propto 1/a$, see \S\ref{sec:discussion}). 
For the $\hra$ disk, the latter process seems to be the dominant one
as reflected by its collisionally-replenished 
{\it secondary} nature -- 
the amount of dust incorporated into larger bodies 
which act as a source of replenishment 
must be much larger than that of the dust lost through 
Poynting-Robertson drag.

\subsection{Predictions for {\it SIRTF}\label{sec:sirtf}}
The {\it Space Infrared Telescope Facility} (SIRTF) will be
capable of sensitive imaging using the {\it Infrared Array Camera} 
(IRAC) at 3.6, 4.5, 5.8, and 8.0$\mum$, and using the 
{\it Multiband Imaging Photometer} (MIPS) at 24, 70, and 160$\mum$.
In Table \ref{tab:sirtf} we show the band-averaged intensities
for our preferred dust models. 

SIRTF will also be able to perform low-resolution 5--40$\mum$
and high-resolution 10--37$\mum$ spectroscopic observations 
using the {\it Infrared Spectrograph} (IRS) instrument.
IR spectroscopy and imaging will provide powerful constraints on 
the $\hra$ dust spatial distribution and its chemical composition.
The ``cold-coagulation'' dust models all predict an almost 
featureless SED (the 9.7$\mum$ amorphous silicate feature 
is very broad and smooth).\footnote{%
 The 3.3, 6.2, 7.7, 8.6 and 11.3$\mum$ vibrational bands 
 characteristic for polycyclic aromatic hydrocarbon (PAH) 
 molecules are not seen in our model spectra although 
 their presence in the $\hra$ disk is expected 
 in the context of the ``cold-coagulation'' model
 since PAHs are a significant constituent of interstellar 
 dust (see e.g. Li \& Draine 2001b). This is mainly because PAHs
 have condensed in the icy mantles of the porous aggregates
 at $r\simgt 70\AU$. In the inner regions of the disk,
 PAHs, sublimated from the icy mantles, would be rapidly 
 photodissociated. Therefore, PAHs are not considered in 
 this paper. The presence of PAHs in dusty disks will
 be discussed in a subsequent paper. 
 }
In contrast, the ``hot-nebula'' dust models predict two sharp 
crystalline silicate features at 11.3 and 23$\mum$.
In order to produce noticeable 11.3 and 23$\mum$ features,
at least 20\% of the silicate dust must be in the crystalline form.
SIRTF spectroscopy will allow us to infer the degree of processing
which the $\hra$ dust has experienced. 

\begin{table}[h]
\caption[]{Dust IR emission (Jy) averaged over SIRTF bands 
           predicted for our preferred models\label{tab:sirtf}.}
\begin{tabular}{cccccccc}
\hline
model & IRAC & IRAC & IRAC & IRAC & MIPS & MIPS &MIPS \\
no.   & 3.6$\mum$ 
      & 4.5$\mum$ 
      & 5.8$\mum$ 
      & 8.0$\mum$ 
      & 24$\mum$
      & 70$\mum$
      & 160$\mum$ \\
\hline
2	&1.18$\times 10^{-8}$
	&9.81$\times 10^{-7}$ 
	&5.09$\times 10^{-5}$
	&7.39$\times 10^{-3}$
	&2.63
	&7.26
	&2.94\\
6	&3.44$\times 10^{-9}$
	&3.91$\times 10^{-7}$ 
	&2.18$\times 10^{-5}$
	&1.67$\times 10^{-3}$
	&2.85
	&6.61
	&2.58\\

13	&1.31$\times 10^{-8}$
	&9.51$\times 10^{-7}$ 
	&4.62$\times 10^{-5}$
	&6.62$\times 10^{-3}$
	&2.56
	&7.81
	&3.41\\
14	&3.68$\times 10^{-9}$
	&3.74$\times 10^{-7}$ 
	&1.98$\times 10^{-5}$
	&1.51$\times 10^{-3}$
	&2.79
	&7.12
	&2.99\\
\hline
\end{tabular}
\end{table}

\section{Conclusion\label{sec:summary}}
We have modelled the mid-infrared to submillimeter spectral 
energy distribution of the ring-like dust disk around the dustiest 
$\sim 8\myr$-old A-type star $\hra$.
We start with two extreme dust types: 
one formed through cold-coagulation of 
unaltered protostellar interstellar grains, 
and one formed through aggregation of grains highly-processed in the 
protostellar nebula with silicate dust annealed and carbon dust oxidized.
We adopt a Gaussian-type dust spatial distribution with a peak at
$70\AU$ from the central star and a FWHM of $15\AU$ as inferred from
images of scattered light and dust thermal emission.
We take a simple power-law for the dust size distribution 
($dn/da\propto a^{-\alpha}$ with $\alpha \approx 2.8-2.9$) 
in the range of $1\mum \simlt a \simlt 1\cm$.
Our principal results are:
\begin{enumerate}
\item It is shown that both types of dust are successful in reproducing
the observed SED, provided that the dust generated by collisions
of planetesimals and cometesimals is highly fluffy, with a vacuum
volume fraction of $\sim 90\%$ (\S\ref{sec:model}).
The fact that the dust in the $\hra$ disk must be somewhere 
intermediate between these two types implies that our models
are robust. Future high resolution spectroscopy of the $\hra$
disk would allow us to infer the degree to which the dust has
been processed (\S\ref{sec:robust}).

\item Although models with a single power-law spatial distribution 
are also able to reproduce the observed SED, the derived dust
distribution is both unphysical and inconsistent with the imaging
observations of scattered light and dust mid-IR thermal emission
(\S\ref{sec:pwl}). 

\item Our models show no evidence for the existence of a hot 
``zodiacal dust'' component suggested to lie in a radial distance 
of a few AU from the star by previous workers to account for 
the emission at $\lambda \simlt 12\mum$ (\S\ref{sec:zodi}). 
Our upper limit on the total mass of this component is at least 
10 times smaller than what would be obtained if the entire 
$\lambda \simlt 12\mum$ emission is attributed to this hot dust.

\item Grains smaller than $\sim 10-20\mum$ will be radiatively 
expelled from the disk; grains at $r=70\AU$ in the range of 
$\sim 10-20\mum \simlt a \simlt 100-200\mum$ will also be removed 
from the disk due to the Poynting-Robertson inward spiralling drag
(at a {\it closer} distance from the star, {\it larger} grains 
will be removed).
Collisions of planetesimals/cometesimals must continuously 
replenish the dust in the disk at a rate of
$\approx 8.3\times 10^{-7}\,m_\oplus/{\rm yr}$ (\S\ref{sec:rppr}).
\end{enumerate}

\acknowledgments
We thank S.E. Strom and the anonymous referee for helpful comments 
and suggestions.
A. Li thanks the University of Arizona for the ``Arizona 
Prize Postdoctoral Fellowship in Theoretical Astrophysics''.
This research was supported in part by grants from the NASA 
origins research and analysis program.

\appendix
\section{Possible Dust Composition for 
Cold Protoplanetary Disk\label{sec:composition}}   
The coagulation of interstellar grains that results in fluffy 
and inhomogeneous aggregates occurs in cold, dense molecular 
clouds and protostellar and protoplanetary dust disks. 
It plays an important role in the formation of 
planetary systems (Weidenschilling \& Cuzzi 1993). 
In this section we approximately derive the proportional 
composition of the dust in circumstellar 
disks around (pre-)MS stars from the abundances of the condensible 
elements (C, N, O, Si, Fe, and Mg),\footnote{%
 Some H will be present, mostly in combination with O, C, and N.
 } 
assuming protostellar activities impose little modification on
protostellar grain compositions (see Beckwith, Henning, \& Nakagawa 2000). 

Let $\xsun$ be the cosmic abundance of X relative to H (we assume 
the cosmic elemental abundances are those of the solar values:
$\csun \approx 391$ parts per million (ppm),
$\nsun \approx 85.2\ppm$, $\osun \approx 545\ppm$,
$\mgsun \approx 34.5\ppm$, $\fesun \approx 34.4\ppm$,
and $\sisun \approx 28.1\ppm$ [Sofia \& Meyer 2001]);
$\xgas$ be the amount of X in gas phase
($\cgas\approx 140\ppm$, $\ngas\approx 61\ppm$, 
$\ogas\approx 310\ppm$; Fe, Mg and Si are highly depleted in dust;
see Li \& Greenberg 1997 and references therein);
$\xdust$ be the amount of X relative H locked up in dust
($\cdust = \csun-\cgas \approx 251\ppm$, $\ndust\approx 24.2\ppm$,
$\odust\approx 235\ppm$, $\mgdust\approx 34.5\ppm$,
$\fedust\approx 34.4\ppm$, $\sidust\approx 28.1\ppm$).
Assuming a stoichiometric composition of MgFeSiO$_4$ for 
interstellar silicates, the total mass of silicate dust per H atom is
$\msil \approx \fedust\mufe + \mgdust\mumg + \sidust\musi + \osil\muo
\approx 5.61\times 10^{-3}\,\muh$ 
where $\mux$ is the atomic weight of X in unit of 
$\muh\approx 1.66\times 10^{-24}\g$,
and $\osil\approx 4\,(\fedust + \mgdust + \sidust)/3\approx 129\ppm$  
is the amount of O in silicate dust per H atom (i.e., we assign 
4 O atoms for the average of the Fe, Mg, and Si abundances).
The carbonaceous dust component is dominated by C, 
with little H, N, and O (we assume H/C=0.5, O/C=0.1). 
The total mass of carbon dust per H atom is 
$\mcarb \approx \cdust\muc + \ndust\mun + 0.5 \cdust\muh + 0.1 \cdust\muo
\approx 3.88\times 10^{-3}\,\muh$. 
The C, O, and N atoms left over after accounting for the silicate
and carbon dust components are assumed to condense in icy grains 
in the form of H$_2$O, NH$_3$, CO, CO$_2$, CH$_3$OH and CH$_4$ 
(following Greenberg [1998], we assume 
CO:CO$_2$:CH$_3$OH:CH$_4$:H$_2$CO=10:4:3:1:1).
The total mass of icy grains per H atom is 
$\mice \approx \mice^{\rm C} + \mice^{\rm N} + \mice^{\rm water}$,
where the mass of C-containing ice $\mice^{\rm C}\approx \cgas\muc 
+ \cgas\,(22\muo+18\muh)/19\approx 2.87\times10^{-3}\,\muh$;
the mass of NH$_3$ ice $\mice^{\rm N}\approx \ngas(\mun+3\muh)
\approx 4.10\times 10^{-4}\,\muh$; 
the mass of water ice $\mice^{\rm water}\approx \owater(\muo+2\muh)
\approx 4.12\times 10^{-3}\,\muh$;
$\owater \approx \osun-\osil-0.1\cdust-22\cgas/19 \approx 229\ppm$ 
is the amount of O locked up in H$_2$O ice (we assume H$_2$O contains
all the remaining available O).

Therefore, as a first approximation, we may assume a mixing ratio
of $\mcarb/\msil \approx 0.7$ and $\mice/(\msil+\mcarb) \approx 0.8$
for cold regions (for hot regions where ices sublimate the dust can be
simply modelled as porous aggregates of silicate and carbon particles with
$\mcarb/\msil \approx 0.7$). This does not deviate much from the in situ 
measurements of cometary dust
($\mcarb/\msil \approx 0.5$, $\mice/[\msil+\mcarb] \approx 1.0$;
see Greenberg \& Li 1999 and references therein) which is often 
suggested to be porous aggregates of unaltered interstellar dust
(Greenberg 1998; Greenberg \& Li 1999).

\section{Dust Morphology\label{sec:porosity}}  
Dust formed through coagulation of many small subgrains
has a porous structure. Let $\vsil$ and $\vcarb$ be the 
total volumes taken up by the silicate and carbonaceous dust 
components, respectively; $P$ be the porosity -- the fractional
volume of vacuum; $\rhosil\,(\approx 3.5\g\cm^{-3})$ 
and $\rhocarb\,(\approx 1.8\g\cm^{-3}$) be the mass densities
of silicate and carbonaceous materials, respectively. 
The mass density for a fluffy aggregate of silicate and carbonaceous
subgrains with a porosity of $P$ is $\langle\rho\rangle = \left(1-P\right) 
\left(\rhosil \vsil + \rhocarb\vcarb\right)/\left(\vsil+\vcarb\right)$.
In regions colder than $\approx 110-120\K$, pre-existing or recondensed
ices around the silicate and carbonaceous dust cores would fill all or part
of the vacuum in the aggregate. Let $\vice$ be the total volume taken up 
by the ice component; $\rhoice\,(\approx 1.2\g\cm^{-3})$ be the ice mass 
density. The porosity for the fluffy aggregate of ice-coated silicate 
and carbonaceous subgrains would be reduced to
$\Pice = \max\left\{0, 1-\left(1-P\right)
\left[1+\vice/\left(\vsil+\vcarb\right)\right]\right\}$.
Its mean mass density is $\langle\rho^{\prime}\rangle = \left(1-\Pice\right) 
\left(\rhosil \vsil + \rhocarb\vcarb + \rhoice\vice\right)/
\left(\vsil+\vcarb+\vice\right)$. The dust ``size'' is defined
as the radius of the sphere which encapsulates the whole fluffy
aggregate. The mean mass densities defined here are independent 
on aggregate size.

\end{document}